\documentclass[twocolumn]{autart}    

\usepackage{graphicx}          
\let\theoremstyle\relax
\usepackage{amsthm}
\usepackage{cite}
\usepackage{amsmath,amssymb,amsfonts}
\usepackage{graphicx}
\usepackage{textcomp}
\newtheorem{assumption}{Assumption}
\newtheorem{corollary}{Corollary}
\newtheorem{theorem}{Theorem}
\newtheorem{remark}{Remark}
\newtheorem{lemma}{Lemma}

\usepackage{appendix}
\usepackage{setspace}
\usepackage{multirow}
\usepackage{xcolor}
\usepackage{bbding}
\usepackage{booktabs}
\usepackage{subfigure}
\usepackage{threeparttable}
\usepackage{bm}
\begin{document}

\begin{frontmatter}

\title{SVR-based Observer Design for Unknown Linear Systems: Complexity and Performance}


\thanks[footnoteinfo]{This paper was not presented at any IFAC
meeting. 
Corresponding author Jianping He.
}

\author[Shangjiao]{Xuda Ding}\ead{dingxuda@sjtu.edu.cn},    
\author[Oxford]{Han Wang}\ead{han.wang@eng.ox.ac.uk},    
\author[Shangjiao]{Jianping He}\ead{jphe@sjtu.edu.cn},    
\author[Shangjiao]{Cailian Chen}\ead{cailianchen@sjtu.edu.cn},  
\author[Shangjiao]{Xinping Guan}\ead{xpguan@sjtu.edu.cn}    

\address[Shangjiao]{The Department of Automation, Shanghai Jiao Tong University, China}  
\address[Oxford]{The Department of Engineering Science, University of Oxford, Oxford, United Kingdom}             


\begin{abstract}                          
In this paper we consider estimating the system parameters and designing stable observer for unknown noisy linear time-invariant (LTI) systems. We propose a Support Vector Regression (SVR) based estimator to provide adjustable asymmetric error interval for estimations. This estimator is capable to trade-off bias-variance of the estimation error by tuning parameter $\gamma > 0$ in the loss function.
This method enjoys the same sample complexity of $\mathcal{O}(1/\sqrt{N})$ as the Ordinary Least Square (OLS) based methods 
but achieves a $\mathcal{O}(1/(\gamma+1))$ smaller variance.
Then, a stable observer gain design procedure based on the estimations is proposed. The observation performance bound based on the estimations is evaluated by the mean square observation error, which is shown to be adjustable by tuning the parameter $\gamma$, thus achieving higher scalability than the OLS methods. The advantages of the estimation error bias-variance trade-off for observer design are also demonstrated through matrix spectrum and observation performance optimality analysis.
Extensive simulation validations are conducted to verify the computed estimation error and performance optimality with different $\gamma$ and noise settings.
The variances of the estimation error and the fluctuations in performance are smaller with a properly-designed parameter $\gamma$ compared with the OLS methods.

\end{abstract}

\end{frontmatter}
\section{Introduction}
\label{sec:introduction}

Parameters of system dynamics are fundamental knowledge for the design of model-based controllers and state observers. 
Estimation procedures use input and output data to estimate system parameters, also called the data-driven modeling methods.
Several methods have been proposed in the past decades.
Conventional system identification methods such as prediction error method  \cite{ljung1999system,abdalmoaty2019linear} and instrumental variable method \cite{ljung1999system,young1970instrumental} are broadly investigated.
Later, subspace model identification (SMI) was proposed to improve the performance of the conventional identification methods and estimate the state-space model directly \cite{chou1997subspace,haber2014subspace,yu2019subspace,inoue2019subspace}. 
Another technique to estimate the system dynamics is the adaptive observer design, which provides estimation error bound and shows promising results under the assumption that the system state matrix is in an observer canonical form \cite{luders1973adaptive,zhang2002adaptive,oliva2016fixed}.
With the development of artificial intelligence and machine learning techniques, artificial neural networks-based modeling \cite{carleo2017solving} and support vector machines-based modeling \cite{cherkassky2004practical} have been proposed.
Recently, thanks to the abilities of self-learning and adaptability, the adaptive dynamic programming-based method based on reinforcement learning (RL) has demonstrated the capability to find the optimal control policy and solve the Bellman equation in a practical way \cite{lewis2009reinforcement,wang2009adaptive,gao2016adaptive,wei2019discrete}.
These methods can obtain near-accurate estimation to some extent, in the sense of asymptotic convergence \cite{bauer2000analysis,ljung1998system}, minimum risks\cite{cherkassky1999model,chalimourda2004experimentally}. Also some work obtained the estimation error bounds under strict assumptions\cite{luders1973adaptive,zhang2002adaptive,oliva2016fixed}.
However, the previous studies only provide asymptotic error bounds when the number of collected samples grows to infinity, and no finite-sample guarantees are available.
More recently, it has been proposed to use finite samplings instead \cite{tsiamis2020sample,dean2020sample,zheng2021sample}. These methods are more practical and beneficial for robust controller design.
\begin{table*}[t]
\caption{Comparisons of the proposed estimator and the existing methods in the literature}
\begin{threeparttable}
\begin{tabular}{ccccccccc}
\toprule[1.5pt]
{Paper} & {Measurement} & {Stability} & {Rollouts} & {Data} & {Complexity} & {Estimator} & {Trade-off\tnote{*}} & {Interval\tnote{**}} \\
\midrule[1pt]
This work              & \multirow{3}{*}{Full}                         & Any                        & Multiple                  & All                   &  \multirow{3}{*}    {$\mathcal{O}(\frac{1}{\sqrt{N}})$}                        & {SVR}                        &   {\checkmark}                         & {Tunable}                           \\\cline{7-9}
Dean, et al.  \cite{dean2020sample}                 &                              & Any                        & Multiple                  & Final                 &                              & \multirow{2}{*}{OLS}                        &   \multirow{2}{*} {$\bm{-}$}                  & \multirow{2}{*}{Fixed}                           \\
Simchowitz, et al.  \cite{simchowitz2018learning}            &                              &           $\rho(\bm{A})<1$                 & Single                    & All                   &                              &                         &                            &                          \\ 
\midrule[0.75pt]
Zheng, et al.    \cite{zheng2021sample}               & \multirow{5}{*}{Partial}     & Any                        & Multiple                  & All                   &     \multirow{5}{*}    {$\mathcal{O}(\frac{1}{\sqrt{N}})$}                               & \multirow{5}{*}{OLS}                        &    \multirow{5}{*} { $\bm{-}$}                         & \multirow{5}{*}{Fixed}                           \\
Sun, et al.    \cite{sun2020finite}                 &                              & Any                        & Multiple                  & Final                 &                              &                         &                            &                            \\
Oymak, et al.   \cite{oymak2021revisiting}                &                              &                $\rho(\bm{A})<1$            & Single                    & All                   &                              &                         &                            &                            \\
Sarkar, et al.    \cite{sarkar2021finite}              &                              &               $\rho(\bm{A})<1$             & Single                    & All                   &                              &                         &                            &                            \\
Simchowitz, et al.    \cite{simchowitz2019learning}          &                              &                   $\rho(\bm{A})<1$         & Single                    & All                   &                              &                         &                            &                            \\ 
\bottomrule[1.5pt]
\end{tabular}
\begin{tablenotes}
        \item[*] The ability to trade-off the bias and variance relationship, and achieve an adjustable estimation.
        \item[**] The estimation error bound interval under a fixed sample length $N$.
\end{tablenotes}   
\end{threeparttable}
\label{tab.1}
\end{table*}

Recent works in the control community discussed non-asymptotic analysis and obtained bound of bias for the dynamic estimation problem\cite{dean2020sample,simchowitz2018learning,zheng2021sample,sun2020finite,oymak2021revisiting,sarkar2021finite,simchowitz2019learning,tsiamis2021linear}.
Based on the OLS estimator, these works considered the Gaussian process estimation of a linear time-invariant (LTI) system. 
The principal tool used in the non-asymptotic regression bound analysis is concentration inequalities.
The convergence rate of regression error is $\mathcal{O}(1/\sqrt N)$, where $N$ is the number of samples (sample length) used for regression.
A line of recent works has obtained similar results for both stable and unstable systems by different data collection procedures.
Based on these results, the robust controller design problem is further considered, which explores the relationships among data, estimation error, and controller performance\cite{zheng2021sample,tsiamis2020sample,dean2020sample}. Specifically, they connected the sub-optimal control cost with the regression error based on the parameterization methods such as system-level synthesis (SLS)\cite{wang2019system}, and input-output parameterization (IOP)\cite{furieri2019input,zheng2020equivalence}.
These works showed that the more data used for OLS regression, the tighter the error bound is and the better control performance. These methods obtained error bounds based on the sample length without considering the impact of variance of the estimation error. 
An ideal estimator should be unbiased and have low variance.
However, such estimators do not exist under finite samples\cite{mehta2019high}. 
Since a larger variance of estimation would cause more uncertainty on regression results, the variance has a great impact on stable observer design.
This motivates us to design an alternative estimator with a smaller variance.

Machine learning methods such as neural networks\cite{geman1992neural} and support vector machines\cite{valentini2004bias} can provide biased but low-variance regression results, which fluctuate moderately.
This characteristic makes the machine learning methods popular in practice.
The bias causes the movement of the error interval, which also influences the robust optimal control formulation.
Besides, trading off the bias and variance provides flexible error bounds, which could benefit the stable controller design.
Therefore, the sample complexity and controller design based on the estimation with bias-variance trade-offs need to be formulated and analyzed.

Since the support vector regression (SVR) techniques have roots in statistical learning theory and promising empirical outcomes in practical applications, they show potential for the sample complexity analysis and the application of the system dynamics estimation.
However, it is not easy to analyze the sample complexity and design a stable observer gain over an unknown system.
The challenges are:
1) the distribution of the estimation based on SVR is unknown, which causes difficulty in analyzing the relationship between the data sample length and the estimation accuracy;
2) the design of the observer gain over the unknown system to guarantee the stability based on the biased system dynamics estimation;
3) hard to formulate the relationship between the designed observer performance and the sample length.
Therefore, this paper introduces SVR techniques into the system dynamics estimation procedure, designs a stable observer gain based on estimation, and formulates the sample complexity of the observer performance.
Specifically, the connection between OLS and SVR is formulated by introducing an $L_2$ loss function into SVR. The distribution of the SVR-based estimation is obtained, and estimation error bounds based on finite data samples are formulated using non-asymptotic analysis. 
We further show that the proposed estimator also has a smaller variance than the OLS-based one, and the error interval of the estimation based on the proposed method can be adjusted when the sample length is fixed. 
This property endows our method with higher scalability beyond the OLS estimator.
Then, the condition of the stable observer is given, and the observer design algorithm is proposed based on estimation results. Moreover, the observer is used to analyze the controller performance. 
The adjustable interval is beneficial for the observer design since a suitable parameter $\gamma$ can introduce less uncertainty, thus helping with stabilizing the system.
Based on the designed observer gain, the performance bound of the observer defined by the mean squared observation error is given based on estimation results.
Furthermore, the performance of the sub-optimal observer is analyzed with SLS techniques.



The main contributions of this paper are threefold:
\begin{itemize}
    \item A SVR-based estimator is proposed to estimate LTI system dynamics with the bias-variance trade-off. To the best of our knowledge, it is the first work on analyzing SVR, a machine learning method, to sample complexity bounds for dynamic estimation and observer design of an unknown linear system.
    The sample complexity of estimation error based on the proposed method reaches $\mathcal{O}(1/\sqrt{N})$.
    Further, the estimation error bound can be adjusted by the bias-variance trade-off with parameter $\gamma$, which provides higher flexibility for stable controller design. Table. \ref{tab.1} shows the comparisons of the proposed method with the existing works.
    \item Based on the dynamic estimation results and the error bound, an observer gain is designed to guarantee stability.
    We show that the adjustable interval helps find a stable gain, which shows that the proposed estimator is beneficial for designing a robust controller for an unknown system.
    \item We further analyze the end-to-end sample complexity of the sub-optimal observer for an unknown system.
    The mean squared observation error bounds are only related to the upper bound of the dynamic estimation error. 
    The result is applicable to both fully and partially observed systems.
\end{itemize}

The paper is organized as follows:
the considered problem setting, the SVR-based estimation procedure, and the observer design goal are shown in Section \ref{sec:2}. 
The non-asymptotic analysis and estimation error bounds based on SVR are discussed in Section \ref{sec:3}.
In Section \ref{sec:design}, we propose a stable observer gain design procedure based on estimation error bounds given in Section \ref{sec:3} for an unknown LTI system.
Then, the performance of the designed observer of the unknown LTI system is formulated in Section \ref{sec:5}.
The numerical simulations are conducted in Section \ref{sec:6}, to show the advantages of the proposed estimator both in dynamic estimation and controller design.
Finally, conclusions and future research directions are given in Section \ref{sec:7}.

\textbf{Notation}. 
We let bold symbols denote the vectors and matrices. $\|\cdot\|$ is the spectral norm. 
The $\mathcal{H}_2$ norm is defined by $\|\Phi\|_{\mathcal{H}_2}^2 \triangleq \sum_{t=0}^\infty\|\Phi_t\|_F^2$, where $\Phi = \sum_{t=0}^\infty\Phi_tz^{-1}$ is the frequency representation of signals and $\|\cdot\|_F$ is the Frobenius norm.
The $\mathcal{H}_\infty$ norm is defined by $\|\Phi\|_{\mathcal{H}_\infty} \triangleq \sup_{\|z\|=1} \|\Phi(z)\|$.
$\frac{1}{z}\mathcal{RH}_\infty$ is the set of real rational stable strictly proper transfer matrices.
Notations $<$, $\leq$, $>$ and $\geq$ are element-wise inequality.
$\mathbb{E}$ and $\mathbb{V}$ stand for the expectation and covariance, respectively.
Table. \ref{tab:test} shows some important definitions used in this paper.
\begin{table}[t]
\small
 \caption{\label{tab:test}Some Important Definitions} 
 \begin{tabular}{cl}
 \toprule 
  Symbol  & Definition  \\ 
  \midrule
  $\bm{A} \in {\mathbb{R}^{n \times n}}$ & the state matrix\\
  $\bm{B} \in {\mathbb{R}^{n \times m}}$ & the input matrix\\
  $\bm{C} \in {\mathbb{R}^{p \times n}}$ & the output matrix\\
  $\bm{\tilde{A}}\in {\mathbb{R}^{n \times n}}$ & the estimation of state matrix\\
  $\bm{\tilde{B}}\in {\mathbb{R}^{n \times m}}$ & the estimation of input matrix\\
  $\Delta\bm{A}\in {\mathbb{R}^{n \times n}}$ & the estimation error of state matrix\\
  $\Delta\bm{B}\in {\mathbb{R}^{n \times m}}$ & the estimation error of input matrix\\
  $\bm{x}_k \in {\mathbb{R}^{n}}$ & the state variable at time $k$\\
  $\bm{y}_k \in {\mathbb{R}^{p}}$ & the output variable at time $k$\\
  $\bm{u}_k \in {\mathbb{R}^{m}}$ & the input variable at time $k$\\
  $\bm{w}_k \sim \mathcal{N}(0,\sigma^2_w \bm{I}_n)$ & the process noise\\
  $\bm{v}_k \sim \mathcal{N}(0,\sigma^2_v \bm{I}_p)$ & the measurement noise\\
  $\bm {e}_k$ & the observer error at time $k$\\
  $\gamma>0$ & the parameter used in SVR\\
  $\bm{L} \in {\mathbb{R}^{n \times p}}$ & the designed observer gain\\
  $\bm{K} \in {\mathbb{R}^{n \times p}}$ & the optimal observer gain\\
  $J$ & the mean square observation error \\
  $T_0$ & the end-time of each roll-out \\
  $N$ & the number of roll-outs\\
  $M$ & the upper bound of $\|\bm{A}\|$ and $\|\bm{B}\|$\\
  \bottomrule 
 \end{tabular} 
\end{table}


\section{Problem Setup}\label{sec:2}
We consider the multiple-input and multiple-output (MIMO)
LTI system
\begin{subequations}\label{LTI-system}
    \begin{align}
        \bm{x}_{k+1} &= \bm{A}\bm{x}_k + \bm{B}\bm{u}_k + \bm{w}_k, \label{LTI-x}\\
         \bm{y}_{k} &= \bm{C}\bm{x}_{k}
         +\bm{v}_k
         , \label{LTI-y}
    \end{align}
\end{subequations}
where $\bm{A} \in {\mathbb{R}^{n \times n}}$, $\bm{B} \in {\mathbb{R}^{n \times m}}$ and $\bm{C} \in {\mathbb{R}^{p \times n}}$ are the 
state, input and output matrices, respectively.
$\bm{x}_k \in {\mathbb{R}^{n}}$,
$\bm{y}_k \in {\mathbb{R}^{p}}$ and
$\bm{u}_k \in {\mathbb{R}^{m}}$ are the state, output and input variables at time $k$, respectively,
and $\bm{w}_k \sim \mathcal{N}(0,\sigma^2_w \bm{I}_n)$, $\bm{v}_k \sim \mathcal{N}(0,\sigma^2_v \bm{I}_p)$ denote the process and measurement noises.
Throughout this paper, we make the following assumption.
\begin{assumption}
The dynamics in \eqref{LTI-x} are unknown.
$\bm A$ is invertible, $(\bm A, \bm B)$ is stabilizable, and matrices are bounded as $\|\bm A\|, \|\bm B\|\le M$ with $M>0$.
\end{assumption}

Assumption 1 defines the LTI system is controllable and observable, which is a standard condition for observer design.
Then, the assumption of state and input matrices being bounded is reasonable.
Similar assumption is also used in \cite{sarkar2021finite,simchowitz2018learning}.

This paper first focuses on SVR-based estimation and its estimation error bound analysis, which is critical for the observer design, especially when the system matrices are unknown.
\begin{figure}[t]
    \centering
    \includegraphics[width=8cm]{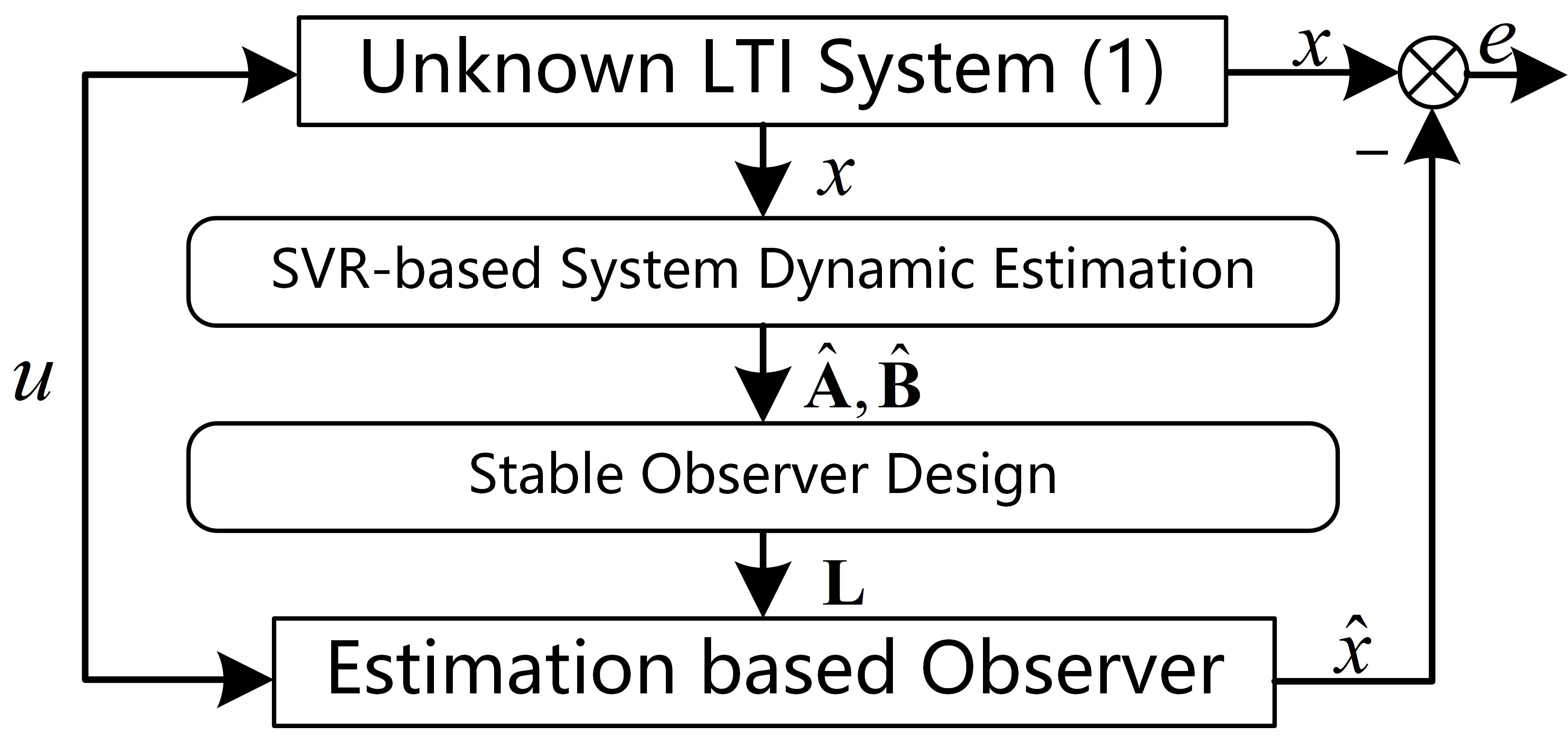}
    \caption{The overview of the proposed estimation and observer synthesis}
    \label{fig:1}
\end{figure}
Our goal is to i) learn the system dynamic matrices $\bm A, \bm B$ with SVR approach, and find the estimation error bound based on a finite number of sample; ii) design a stable observer according to the estimations (denoted by $\tilde{\bm A}, \tilde{\bm B}$) of system matrices; iii) analyze the observer performance and derive the performance bound.
The scheme of the proposed estimation and observer synthesis is shown in Fig. \ref{fig:1}.

\subsection{Data Collection}
Since we do not assume that the system \eqref{LTI-system} is open-loop stable, the state variable $\bm{x}$ might blow up during the data collection process.
Inspired by previous study on OLS regression\cite{    dean2020sample,simchowitz2018learning,zheng2021sample,sun2020finite,oymak2021revisiting,sarkar2021finite,simchowitz2019learning}, we use multi-roll-out procedure to collect $\bm{x}_k$ and $\bm{u}_k$.
The essence of this procedure is to use collected data in a finite time horizon.
The system with $\bm{x}_0 = 0$ is excited by Gaussian input $\bm{u}_k \sim \mathcal{N}(0,\sigma^2_u \bm{I}_m)$.
The data set is then recorded as
\begin{equation*}
    \left \{\left(\bm{x}_k^i, \bm{u}_k^i\right): 1\leq i\leq N, 0\leq k\leq T_0 \right \},
\end{equation*}
where $i$ is the index for each roll-out and $T_0$ is the end time of the roll-out.
The total number of data points is $NT_0$. Gaussian input satisfies the sufficient persistency of excitation condition for linear systems \cite{willems2005note}.

A similar roll-out procedure is used in \cite{sun2020finite,dean2020sample,zheng2021sample} to deal with unstable single-input single-output (SISO) and MIMO systems.
Besides multiple roll-out procedures, single roll-out procedures are used to identify open-loop stable systems \cite{simchowitz2018learning,oymak2021revisiting,sarkar2021finite,simchowitz2019learning}.

\begin{remark}
    $\bm{x}_0$ can follow a Gaussian distribution with finite variance, which does not influence the regression procedure and regression error analysis\cite{zheng2021sample}.
\end{remark}

\subsection{SVR Procedure}
After data collection, SVR-based estimator is used to estimate the system dynamics.
Input and output data are put into tuples for each roll-out,
\begin{subequations}
\begin{align}\nonumber
    f({\bm{z}_k})&= \left [\bm{x}_{k}^1,~ \bm{x}_{k}^2,\cdots,~ \bm{x}_{k}^N \right] \in {\mathbb{R}^{n \times N}},\\ \nonumber
    {\bm{z}_k}&=  \left [
    \begin{aligned}
    \bm{x}_{k-1}^1\\\nonumber
    \bm{u}_{k-1}^1
    \end{aligned},~
    \begin{aligned}
    \bm{x}_{k-1}^2\\\nonumber
    \bm{u}_{k-1}^2
    \end{aligned},
    \cdots,~
    \begin{aligned}
    \bm{x}_{k-1}^N\\
    \bm{u}_{k-1}^N
    \end{aligned}
    \right ] \in {\mathbb{R}^{(n+m) \times N}},
\end{align}
\end{subequations}
and the noise $\bm{w}$ is represented as 
\begin{equation*}
    {\bm{b}_k}=\left [\bm{w}_{k-1}^1,~ \bm{w}_{k-1}^2,\cdots,~ \bm{w}_{k-1}^N \right] \in {\mathbb{R}^{n \times N}}.
\end{equation*}
Combine all the data as 
\[\mathcal{D}_k = [f({\bm{z}_k});{\bm{z}_k};{\bm{b}_k}] \in {\mathbb{R}^{(3n+m) \times N}}.\]
When $k=T_0$, the final data is used to estimate system dynamics, and such procedure is called final-data estimation \cite{dean2020sample,sun2020finite}.
The total number of data in the final-data collection is $(3n+m) \times N$. 
To efficiently use the data collected in the collection procedure, data from $k=1$ to $k=T_0$ are all used for estimation, as $\{\mathcal{D}_1, \mathcal{D}_2,\dots,\mathcal{D}_{T_0}\} \in \mathbb{R}^{(3n+m) \times (T_0-1)N}$.
This procedure is referred to as all-data estimation.

Unlike OLS, which can estimate the coefficients of a matrix at one time, ordinary SVR can only regress one row of a matrix. To ease the notation, we consider the estimation based on $\mathcal{D}_{T_0}$ in the following subsection.
Then, regression of \eqref{LTI-x} for each row of $\left[\bm{A}, \;\bm{B} \right]$ is shown as
\begin{equation*}
     f_i({\bm{z}_k}) = \left[\bm{A}, \;\bm{B} \right]_i{\bm{z}_k} + {\bm{b}}_{k,i},
\end{equation*}
where subscript $i$ stands for the $i$-th row of the matrix, i.e., $f_i({\bm{z}}_k) \in {\mathbb{R}^{1 \times N}}$, $\left[\bm{A}, \;\bm{B} \right]_i  \in {\mathbb{R}^{1 \times (n+m)}}$,
${\bm{b}}_{k,i} \in {\mathbb{R}^{1 \times N}}$.

To minimize the estimation error, regression procedure is formulated with a slight modification on the OLS method.
The regression procedure estimates $\left[ {\bm{A}}, \;  {\bm{B}}\right]_i$ by minimizing
\begin{equation}\label{eq:ridge}
    \frac{1}{2}\left\|\left[ {\bm{A}}, \;  {\bm{B}}\right]_i\right\|^2 +\frac{1}{2\gamma}\sum_{j=1}^N(f_i(\bm{z}_{k}^{j}) - \left[\bm{A}, \; \bm{B} \right]_i{\bm{z}_{k}^{j}})^2,
\end{equation}
where $\bm{z}_{k}^{j} \in \mathbb{R}^{(n+m)\times1}$ is the $j$-th column of $\bm{z}_{k}$, $f_i(\bm{z}_{k}^{j}) \in \mathbb{R}$.
Then, based on Vapnik's theorem \cite{vapnik1999nature}, we formulate the following SVR-based optimizaiton problem.  
\begin{subequations}\label{eq:SVR}
\begin{align}
    \min_{[ {\bm{A}}, \; {\bm{B}}]_i}\label{eq:SVR-1}\mathcal{L}_i = 
	& \frac{1}{2}\left\|\left[ {\bm{A}}, \;  {\bm{B}}\right]_i\right\|^2  +\frac{1}{2\gamma}\sum_{j=1}^N(\xi^{+2}_{j}+\xi^{-2}_{j}),\\
	    \text{s.t.}  ~~&f_i(\bm{z}_{k}^{j}) - \left[\bm{A}, \; \bm{B} \right]_i{\bm{z}_{k}^{j}} - \xi_{j}^+ \leq 0,\\
        & -f_i(\bm{z}_{k}^{j}) + \left[\bm{A}, \; \bm{B} \right]_i{\bm{z}_{k}^{j}} - \xi_{j}^- \leq 0,\\
        &\xi_j^+, \xi_j^- \ge 0 ,
\end{align}
\end{subequations}
where $\xi^+_{j}$ and $\xi^-_{j}$ are slack variables to separate ${\bm{b}}_{k,i}^{j}\geq0$ and ${\bm{b}}_{k,i}^{j}\leq 0$. $\gamma>0$ is the parameter. Although the estimation results cannot be identical to the actual value due to the noise, the estimation goal in this paper is to get a bounded deviation based finite samples.
The estimations of $\bm{A}$ and $\bm{B}$ are given based on solving the dual problem of \eqref{eq:SVR} by introducing dual parameter $\bm{\alpha}\in {\mathbb{R}^{N \times 1}}$ as
\begin{equation}\label{eq:SVM-solve-1}
    \left[\tilde {\bm{A}},\; {\tilde{\bm{B}}} \right]_i = \bm{\alpha}^\top \bm{z}_k^\top,
\end{equation}
where $\tilde{\bm{A}}$ and $\tilde{\bm{B}}$ are the estimations of ${\bm{A}}$ and ${\bm{B}}$, respectively. 
Solving procedure details of \eqref{eq:SVM-solve-1} are shown in Appendix \ref{app:1}. 
The augmented SVR is formulated to estimate $\bm{A}$ and $\bm{B}$, such that
\begin{equation}\label{eq:SVR_all}
    {\min_{[ {\bm{A}} \; {\bm{B}}]}} \sum\limits_{i = 1}^n \mathcal{L}_i .
\end{equation}
Note that the parameter $\gamma$ can be used to trade off bias and variance of the estimation\cite{valentini2004bias}.
When $\gamma \to 0$, the second term of \eqref{eq:SVR-1} dominates the minimization, and the estimation results tend to be the same as the unbiased results of OLS.
The variance of the estimation decreases as parameter $\gamma$ increases.
These methods attempt to decrease the regression variance and improve the model's robustness by sacrificing the unbiasedness property.
However, OLS seeks to obtain an unbiased estimation and overlooks the variance, which could lead to poor performance (i.e., large variance) in regression.
Thus, compared to OLS, we can use SVR with a tunable parameter $\gamma$ to control the tradeoff between bias and variance and improve the estimation performance. 
\begin{remark}
The SVR estimator in this paper is different from $\epsilon$-insensitive SVR.
By using dual parameter $\bm{\alpha}$, the support vectors are introduced into the estimation results.
Thus, the proposed method is an SVR-based one.
Furthermore, by introducing Vapnik's theorem in \eqref{eq:ridge}, the derivation gives extra insight as \eqref{eq:SVM-solve-1}.
The connection between \eqref{eq:ridge} and \eqref{eq:SVR} is broadly studied, we refer the reader to \cite{wahba1999support,saunders1998ridge}.
\end{remark}


\subsection{Observer Synthesis}
After estimating system dynamics based on the data collection and SVR procedure, we focus on the observer gain $\bm L$ design.
The observer that designed based on estimation results $\tilde{\bm{A}}$ and $\tilde{\bm{B}}$ is shown as
\begin{equation*}
    \tilde{\bm{x}}_{k+1} = \tilde{\bm{A}}\tilde{\bm{x}}_{k}+\tilde{\bm{B}}{\bm{u}}_{k} + \bm{L}(\bm{y}_k-\bm{C}\tilde{\bm{x}}_{k}).
\end{equation*}
Then, the observer error $\bm {e}_k = \tilde{\bm {x}}_k - {\bm {x}}_k$ can be written as
\begin{equation}\label{x_tildex}
    \bm {e}_{k+1} =  ({\bm{A}} -  {\bm{LC}}){\bm{e}}_k - \Delta {\bm{A}} \tilde{\bm{x}}_k - \Delta {\bm{B}} \bm{u}_k -  {\bm{w}_k} + \bm{Lv}_k,
\end{equation}
where $\tilde {\bm{A}} = \bm{A} - \Delta \bm{A}$ and $\tilde {\bm{B}} = \bm{B} - \Delta \bm{B}$, $\bm{L} \in {\mathbb{R}^{n \times p}}$ is the observer gain, which is also used in Kalman filter and linear quadratic Gaussian control design.
$\tilde{\bm{x}}_k \in {\mathbb{R}^{n}}$ is the state given by the observer.

Apparently, $\bm{L}$ needs to be designed so that ${\bm{A}} -  {\bm{LC}}$ is stable.
When ${\bm{A}}$ is known, it is easy to design a suitable $\bm{L}$ and make the spectrum radius $\rho({\bm{A}} -  {\bm{LC}}) < 1$.
By referring to \cite{tsiamis2020sample}, we use the mean square observation error $J$ to measure the observer performance,
\begin{equation}\label{eq:observer_ref}
    J = \lim_{T\to\infty}\mathbb{E}\left (\frac{1}{T}\sum_{k=0}^T\| \bm {e}_{k+1}\|^2\right).
\end{equation}

The value of $J$ depends on the dynamics estimation results and designed observer gain.
The two-step procedure, estimation and observer application, give the end-to-end performance analysis\cite{tsiamis2020sample,zheng2021sample,dean2020sample}, which shows that the cost depends on the data sampled in $N$ roll-outs.
When $N \to \infty$, $\|\Delta {\bm{A}}\| \to 0$ and $\| \Delta {\bm{B}}\| \to 0$, and optimal $ {\bm{L}}$ can be designed to make  $\rho({\bm{A}} -  {\bm{LC}}) < 1$
and the cost is 0\cite{alazard2005introduction}.

\section{SVR Estimation Error Bound Analysis}\label{sec:3}
In this section, we show the sample complexity of SVR-based estimation error.
We first show the distribution of SVR estimation for an LTI Gaussian system. 
Define the matrices $\bm{G}_{k}$ and $\bm{F}_{k}$ as
\[ \bm{G}_{k} = \left[ \bm{A}^{k-1}\bm{B} \,\,  \bm{A}^{k-2}\bm{B} \,\, \dots \,\, \bm{B}   \right],\]
\[ \bm{F}_{k} = \left[ \bm{A}^{k-1} \,\,  \bm{A}^{k-2} \,\, \dots \,\, \bm{I}_n   \right].\]
Then, $f\left (\left [
    \begin{aligned}
    \bm{x}_{k-1}\\\nonumber
    \bm{u}_{k-1}
    \end{aligned}
    \right ]\right)$ and $\left [
    \begin{aligned}
    \bm{x}_{k-1}\\\nonumber
    \bm{u}_{k-1}
    \end{aligned}
    \right ]$ follow
\begin{subequations}
\begin{align} \nonumber
    &f\left (\left [
    \begin{aligned}
    \bm{x}_{k-1}\\\nonumber
    \bm{u}_{k-1}
    \end{aligned}
    \right ]\right) \sim \mathcal{N} (0, \bm{\sigma}^2_{k}),\\
    &\left [
    \begin{aligned}
    \bm{x}_{k-1}\\\nonumber
    \bm{u}_{k-1}
    \end{aligned}
    \right ] \sim \mathcal{N} \left( 0, 
    \begin{bmatrix}
    \bm{\sigma}^2_{k-1} & 0\\
    0&\sigma^2_u\bm{I}_{m}
    \end{bmatrix} \right ),\nonumber
\end{align}
\end{subequations}
where $\bm{\sigma}^2_{k} = \sigma^2_u\bm{G}_{k}\bm{G}_{k}^\top + \sigma^2_w\bm{F}_{k}\bm{F}_{k}^\top$.
Notice that the noise will propagate throughout the system with time, and the accumulation is directly reflected  by convolution, which is represented by $\bm{G}_{k}$ and $\bm{F}_{k}$.
By Assumption 1, we have $\|\bm{A}\| \le M$ and $\|\bm B\|\le M$, thus $\mathbb{V}(\bm{Bu}_k)\leq mM\sigma^2_u\bm{I}_n$ and $\bm{\sigma}^2_{k}\leq (nM^{2T_0-1}\sigma^2_u + M^{2T_0-2} \sigma^2_w)\bm{I}_n$ when $k=T_0$.
Then, the regression procedure can be considered into two parts: 
i) regressing $\bm{A}$ with the residual terms $\bm{Bu}_{T_0}$ and $\bm{w}_{T_0}$, which can be treated as the noise together.
Since $\mathbb{V}(\bm{Bu}_k)\leq mM\sigma^2_u\bm{I}_n$ and $\mathbb{V}(\bm{w}_{T_0}) = \sigma^2_w\bm{I}_{n}$, the covariance of the noise when regressing $\bm{A}$ is equal or less than ${\sigma}^2_A\bm{I}_n = (mM\sigma^2_u + \sigma^2_w)\bm{I}_{n}$.
ii) regressing $\bm{B}$ with the residual terms $\bm{Ax}_{T_0}$ and $\bm{w}_{T_0}$, which can be treated as the noise together.
Since $\bm{\sigma}^2_{T_0}\leq (nM^{2T_0-1}\sigma^2_u + M^{2T_0-2} \sigma^2_w)\bm{I}_n$ and $\mathbb{V}(\bm{w}_{T_0}) = \sigma^2_w\bm{I}_{n}$, the covariance of the noise when regressing $\bm{B}$ is equal or less than ${\sigma}^2_B\bm{I}_n = (nM^{2T_0-1}\sigma^2_u + M^{2T_0-2} \sigma^2_w + \sigma^2_w)\bm{I}_{n}$.



\begin{lemma}\label{lemma:1}
    For a linear system with independent Gaussian noise \eqref{LTI-x}, the expectation and covariance of $\tilde {\bm{A}}_i$ estimated by SVR with $L_2$  loss \eqref{eq:SVR_all} are
    \begin{subequations}\label{eq:distribution-A}
    \begin{align}
            \mathbb{E}(\tilde {\bm{A}}_i) &= \frac{1}{1+\gamma} {\bm{A}}_i,\label{eq:distribution-EA}\\
            \mathbb{V}(\tilde {\bm{A}}_i) &\leq \frac{1}{1+\gamma} ({\sigma}^2_A (\bm{x}_{T_0-1}\bm{x}_{T_0-1}^\top)^{-1} +\gamma M^2\bm{I}_n),\label{eq:distribution-CA}
    \end{align}
    \end{subequations}
    and the expectation and the covariance of $\tilde {\bm{B}}_i$
    \begin{subequations}\label{eq:distribution-B}
    \begin{align}
            \mathbb{E}(\tilde {\bm{B}}_i) &= \frac{1}{1+\gamma} {\bm{B}}_i,\label{eq:distribution-EB}\\
            \mathbb{V}(\tilde {\bm{B}}_i) &\leq \frac{1}{1+\gamma} ({\sigma}^2_B (\bm{u}_{T_0-1}\bm{u}_{T_0-1}^\top)^{-1} +\gamma M^2\bm{I}_m),\label{eq:distribution-CB}
    \end{align}
    \end{subequations}
    where $\bm{x}_{T_0-1}$ and $\bm{u}_{T_0-1}$ are deterministic quantities of the samplings.
\end{lemma}
\begin{proof}
Please see Appendix B.
\end{proof}
Note that ${\sigma}^2_A (\bm{x}_{T_0-1}\bm{x}_{T_0-1}^\top)^{-1}$ is the estimation variance based on OLS, thus SVR has a $\mathcal{O}(1/(1+\gamma))$ minor estimation error variance under more considerable variance noise (see Appendix \ref{proof:lemma1}).
Thus, SVR can provide more stable results than OLS when estimating a system with large noise.
Since there is linear relationship between OLS-based estimation and SVR-based one, the estimation results of SVR follow a normal distribution.

The distribution of $\tilde {\bm{A}}_i$ and $\tilde {\bm{B}}_i$ cannot be used to quantify the error bound with a certain value.
Then, Theorem 1 is given to quantify the estimation error bound.
\begin{theorem}\label{tho:sample-complexity}
Given a constant $\delta$  (where $0<\delta<1$). If all-data in $N$ roll-outs from beginning to $T_0$ is used for estimation based on SVR in \eqref{eq:SVR_all}, then we have the bounds with probability at least $1-\delta$ as,
    \begin{equation}\label{eq:16}
    \begin{aligned}
        &\frac{1}{1+\gamma}\|\gamma {\tilde{\bm{A}}_i - \Delta {\bm{A}}_i} \|\\
        &\leq \sqrt{\frac{\theta_A +n\gamma M^2}{(1+\gamma)N_0}}+ \sqrt{\frac{2(\theta_A+n\gamma M^2)\log(1/\delta)}{(1+\gamma)N_0}},
    \end{aligned}
    \end{equation}
    \begin{equation}\label{eq:17}
    \begin{aligned}
        &\frac{1}{1+\gamma}\|\gamma {\tilde{\bm{B}}_i - \Delta {\bm{B}}_i} \|\\
        &\leq \sqrt{\frac{\theta_B +m\gamma M^2}{(1+\gamma)N_0}}+ \sqrt{\frac{2(\theta_B+m\gamma M^2)\log(1/\delta)}{(1+\gamma)N_0}},\\
    \end{aligned}
    \end{equation}
where $N_0 = (T_0-1)N$, $\theta_A = \frac{4n(mM\sigma^2_u + \sigma^2_w)}{N(nM^{2T_0-1}\sigma^2_u + M^{2T_0-2} \sigma^2_w)}$ and $\theta_B =  \frac{4m(nM^{2T_0-1}\sigma^2_u + (M^{2T_0-2}+1) \sigma^2_w)}{N\sigma^2_u}$.
\end{theorem}
\begin{proof}
Please see Appendix C.
\end{proof}
Theorem \ref{tho:sample-complexity} formulates the relationship between the estimation and the estimation error.
Different from the sample complexity of OLS-based estimator, which directly gives the bound of the estimation error, theorem \ref{tho:sample-complexity} shows the bound of $\|\gamma {\tilde{\bm{A}}_i - \Delta {\bm{A}}_i} \|$ for the bias-variance trade-off in the estimation.
Since $\tilde{\bm{A}}_i$ can be obtained when the training sample and sample length are given, the uncertainty of estimation error is quantified in theorem \ref{tho:sample-complexity}.
Theorem \ref{tho:sample-complexity} also states that the sample-complexities of $\Delta {\bm{A}}_i$ and $\Delta {\bm{B}}_i$ behavior as $\mathcal{O}(1/\sqrt{N})$, which is consistent with the previous studies based on OLS estimation\cite{simchowitz2018learning,oymak2021revisiting,sarkar2021finite,zheng2021sample,dean2020sample,sun2020finite}.
Further, the parameter $\gamma$ can change the bound of estimation error under fixed sample length $N$.
The essence lies in the bias-variance trade-off based on the parameter $\gamma$ in SVR.
Then, the interval of the estimation error is defined as $\mathcal{B}(c,r) = [c-r,c+r]$, where $c$ is the center, $c-r$ and $c+r$ are bounds.
\begin{lemma}\label{lemma:interval}
    Suppose the condition in Theorem \ref{tho:sample-complexity} holds, then the estimation error of the elements of $\bm{A}$ and $\bm{B}$ are in the intervals with probability at least $1-\delta$,
    \begin{subequations}\label{eq:18}
    \begin{align}
        \Delta {\bm{A}_{i,j}} \in 
        \mathcal{B}\left(\gamma\tilde{\bm{A}}_{i,j},\sqrt{(1+\gamma)\mathcal{H}_A} \right),\label{eq:18a}\\
        \Delta {\bm{B}_{i,j}} \in 
        \mathcal{B}\left(\gamma\tilde{\bm{B}}_{i,j},\sqrt{(1+\gamma)\mathcal{H}_B} \right),\label{eq:18b}
    \end{align}
    \end{subequations}
    where $\mathcal{H}_A = \sqrt{\frac{\theta_A +n\gamma M^2}{(1+\gamma)N_0}}+ \sqrt{\frac{2(\theta_A+n\gamma M^2)\log(1/\delta)}{(1+\gamma)N_0}}$,
    and 
    $\mathcal{H}_B = \sqrt{\frac{\theta_B +m\gamma M^2}{(1+\gamma)N_0}}+ \sqrt{\frac{2(\theta_B+m\gamma M^2)\log(1/\delta)}{(1+\gamma)N_0}}$.
\end{lemma}
\begin{proof}
Please see Appendix D.
\end{proof}

	Lemma \ref{lemma:interval} further illustrates that $\gamma$ can change the interval of error.
	The bias-variance trade-off lies in the different variation rates of the center and the radius in \eqref{eq:18}.
	Note that Lemma \ref{lemma:interval} does not bound the OLS-based estimation result tightly due to the inequalities used in Lemma \ref{lemma:1}.
	More importantly, $\gamma$ can change the bound of the error interval, which benefits the uncertainty analysis.
	Take the estimation of $\bm{A}$ as an example.
	It is easy to derive that interval of $\bm{A}$ can be obtained by the estimation procedure with probability at least $1-\delta$ as
\begin{equation}\label{eq:inter:A}
        {\bm{A}_{i,j}} \in 
        \mathcal{B}\left((1+\gamma)\tilde{\bm{A}}_{i,j},\sqrt{(1+\gamma)\mathcal{H}_A} \right).
\end{equation}

	When analyzing the stability of $\bm{A}$, the spectral radius needs to be examined.
	Notice that the expectation of SVR-based estimation is $1/(1+\gamma)$ less than the actual value.
	The biased estimation in \eqref{eq:inter:A} provides a way to adjust the asymmetric interval of $\bm{A}_{i,j}$ by tuning $\gamma$.
	
	Bounds of $||\Delta \bm{A}||$ and $||\Delta \bm{B}||$ can be obtained by the following corollary.
	\begin{corollary}\label{co:1}
Suppose Theorem 1 holds true, then $||\Delta \bm{A}||\leq \epsilon_A$ and $||\Delta \bm{B}||\leq \epsilon_B$, where $\epsilon_A$ and $\epsilon_B$ are defined by
\begin{subequations}\label{eq:norm_bound}
\begin{align}
    \epsilon_A := \sqrt{\sum \limits_{i=1}^{n}\sum\limits_{j=1}^{n}\left(||\gamma\tilde{\bm{A}}_{i,j}||+\sqrt{(1+\gamma) \mathcal{H}_A}\right)^2},\\
    \epsilon_B := \sqrt{\sum \limits_{i=1}^{n}\sum\limits_{j=1}^{n}\left(||\gamma\tilde{\bm{B}}_{i,j}||+\sqrt{(1+\gamma) \mathcal{H}_B}\right)^2}.
\end{align}
\end{subequations}
\end{corollary}
\section{Results on Stable Gain Design}\label{sec:design}
After estimation and obtaining the error interval, the observer gain $\bm{L}$ needs to be designed to guarantee the stability of the observer. 
Specifically, the goal is to determine $\bm{L}$, so that $\bm{A-LC}$ is stable.
However, only the interval of $\bm{A}$ can be obtained by the estimation procedure as \eqref{eq:inter:A}.
Then $\bm{A-LC}$ is in the interval with probability at least $1-\delta$ as
\begin{equation}\label{eq:inter:A-LC}
\begin{aligned}
    &{\bm{A}_{i,j}}- \{\bm{LC}\}_{i,j} \in\\
        &\mathcal{B}\left((1+\gamma)\tilde{\bm{A}}_{i,j}- \{\bm{LC}\}_{i,j},\sqrt{(1+\gamma)\mathcal{H}_A} \right),
\end{aligned}
\end{equation}
where $\{\bm{LC}\}_{i,j}$ is the $i,j$-th element in $\bm{LC}$.
Then, the design of stable observer gain is equivalent to finding a suitable $\bm{L}$, such that the spectral radius of the matrix $\bm{A}-\bm{LC}$ is smaller than $1$.
Determining the robustness of a given observer gain (controller) under uncertainty is widely studied.
Several methods are developed to infer the range of uncertainty for the robust controller \cite{balas1993mu,hjartarson2015lpvtools,doyle1989state}.
However, designing a robust observer gain (controller) is still a hard problem \cite{scherer2001theory}.
A conservative design for a stable observer gain is given in Theorem \ref{th:2}, where $\bm{L}$ guarantees that $\bm{A} - \bm{LC}$ is stable for all combination in \eqref{eq:inter:A-LC}.
	\begin{theorem}\label{th:2}
		Suppose (\ref{eq:inter:A-LC}) holds true, when observer gain $\bm{L}$ satisfies 
		\begin{subequations}
		\begin{align}
		 &\|{\bm{A}_{i,i}}- \{\bm{LC}\}_{i,j}\|<1,\label{eq:choose1}\\
		\sum\limits_{i \ne j,j = 1}^{j = n}     	&\left \{\|(1+\gamma){\bm{A}_{i,j}}- {\{\bm{LC}\}_{i,j}}\|+\sqrt{(1+\gamma)\mathcal{H}_A}\right\}\label{eq:choose2}\\
    	&\leq 1-\|{\bm{A}_{i,i}}-	\{\bm{LC}\}_{i,j}\|.\nonumber
		\end{align}
		\label{eq:choose}
		\end{subequations}
Then, the observer with $\bm{L}$ is stable with probability at least $1-\delta$. 
	\end{theorem}
	\begin{proof}
Please see Appendix E.
\end{proof}
	\begin{figure}[t]
    \centering
    \includegraphics[width=8.5cm]{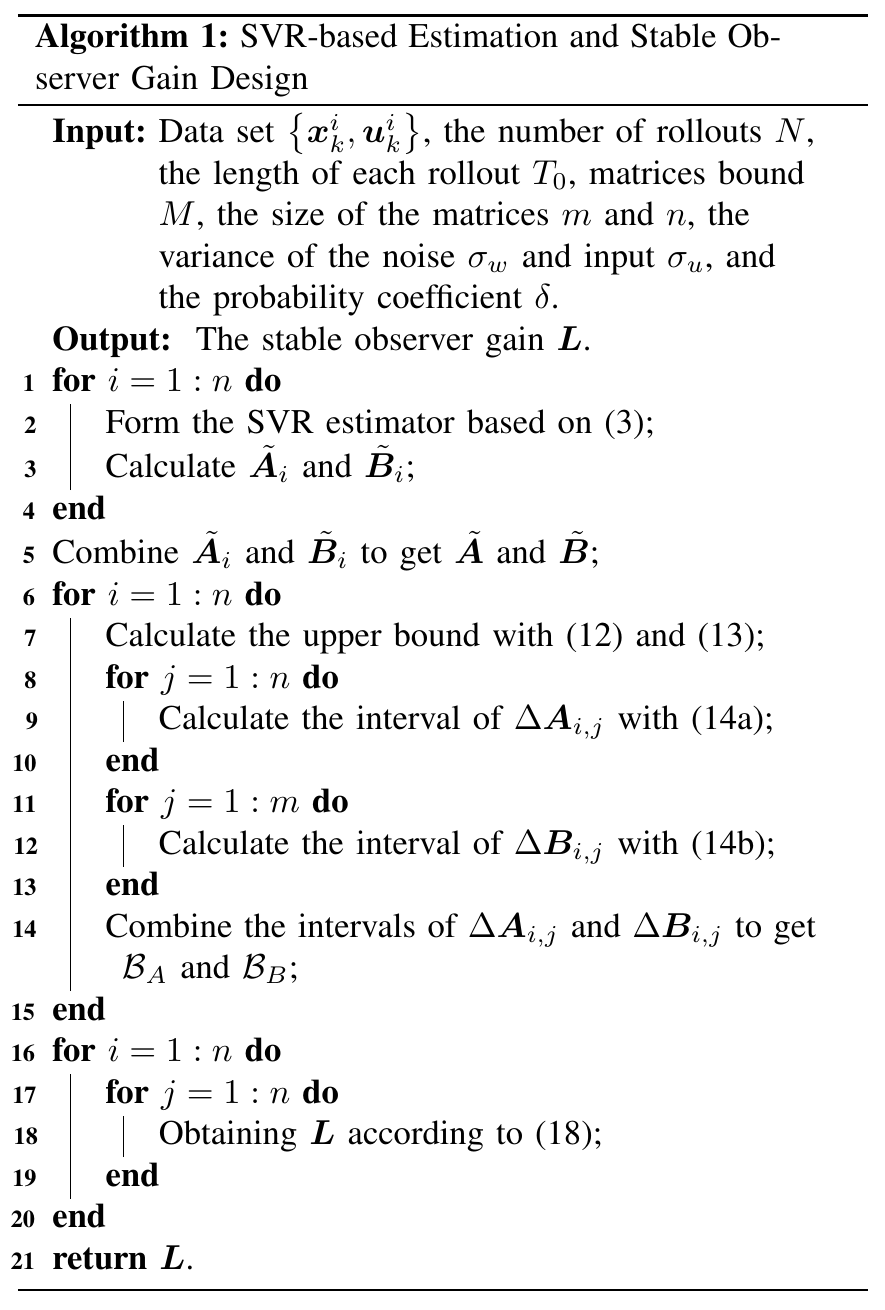}
\end{figure}
Theorem \ref{th:2} is valid for a general form of matrix $\bm{A}-\bm{LC}$.
It should be noted that the observer gain can be easily designed when the uncertain system is with weakly-coupled states, i.e., the norm of diagonal element is much larger than that of other element.
Algorithm 1 shows the procedure of estimation and observer gain design.
\begin{remark}
\eqref{eq:choose1} shows that a smaller range of interval of \eqref{eq:inter:A} leads to finding $\bm{L}$ more easily.
Fortunately, we can trade off the bias-variance in SVR by tuning $\gamma\to0$ in \eqref{eq:SVR_all} and further obtain the smaller range of interval.
However, to satisfy the inequality in \eqref{eq:choose2}, the sum term needs to be small.
Note that the error bound is asymmetric. 
The center of the error bound is away from $0$ as $\gamma$ increases, leading to the sum term decreasing.
Therefore, there is a trade-off in bias-variance for designing a stable observer gain.
\end{remark}
We illustrate the influence of parameter $\gamma$ to observer design in numerical simulations in Section \ref{sec:6}.

\section{Results on Sub-Optimal Observer Analysis}\label{sec:5}
The observer performance is first analyzed based on the observer gain designed in Section \ref{sec:design}.
Then, a sub-optimal observer is constructed based on System Level Synthesis(SLS) framework \cite{wang2019system}.

First, suppose there exists an optimal observer gain $\bm{K}$, which is obtained by solving the Ricatti equation of the observer \cite{kailath2000linear} and $J = 0$ in \eqref{eq:observer_ref}.
When the system dynamic is known, the observer error $\bm{e}(z)$ in \eqref{x_tildex} is given with $z$-transfer into frequency domain according to (5) of \cite{tsiamis2020sample} as
\begin{equation}\label{eq:error_new}
    (z\bm{I-A+LC})^{-1}\bm{Kv}-(z\bm{I-A+LC})^{-1}\bm{Lv}.
\end{equation}
Notice that only the measurement noise $\bm{v}$ displays in \eqref{eq:error_new} since the observer gain $\bm{K}$ has already balanced the process noise $\bm{w}$ and measurement noise $\bm{v}$.
Define the noise $\bm{e}(z)$ responses to $\bm{Kv}$ and $\bm{v}$ by $\Phi_w \triangleq (z\bm{I-A+LC})^{-1}$ and $\Phi_v \triangleq (z\bm{I-A+LC})^{-1}\bm{L}$, respectively.
Then, the observer error $\bm{e}(z)$ is given as
\begin{equation*}
    \bm{e} = ({\Phi_w\bm{K}- \Phi_v})\bm{v},
\end{equation*}
$\bm{L}$ is parameterized as $\Phi_w^{-1}\Phi_v$, where the close-loop responses $\Phi_w$ and $\Phi_v$ are in the set of real rational stable strictly proper transfer matrices $\frac{1}{z}\mathcal{RH}_\infty$.
When the observer is internally stable, the mean squared error is 
\begin{equation*}
    J = \|({\Phi_w\bm{K}- \Phi_v})\sigma_v\|_{\mathcal{H}_2},
\end{equation*}
where $\|\cdot\|_{\mathcal{H}_2}$ is the ${\mathcal{H}_2}$ norm. When $\bm{L} = \bm{K}$, the error-free observer is achieved, and $J=0$.
The observer error $\bm{e}(z)$ is given as
\begin{equation*}
\begin{aligned}
     \bm{e} &=  \tilde{\Phi}_w \Delta \bm{A} \Phi_A \bm{B} \bm{u} + \tilde{\Phi}_w \Delta \bm{A} \Phi_A \bm{K} \bm{v}\\
    &+ \tilde{\Phi}_w \Delta \bm{B} \bm{u} +
    ({\tilde{\Phi}_w\bm{K}- \tilde{\Phi}_v})\bm{v},
\end{aligned}
\end{equation*}
where ${\tilde{\Phi}_w}\triangleq (z\bm{I} -\tilde{\bm{A}}+\bm{LC})^{-1}$,  ${\tilde{\Phi}_v}\triangleq (z\bm{I} -\tilde{\bm{A}}+\bm{LC})^{-1}\bm{L}$, $\Phi_A \triangleq (z\bm{I} -\bm{A})^{-1} $ and $(z\bm{I} -\bm{A})$ is inevitable.
Then, the mean squared error of the state for the observer with uncertainty is given in Lemma \ref{lemma:3}.
\begin{lemma}\label{lemma:3}
    Consider system \eqref{LTI-system} with unknown dynamics. The stable observer with parameterization has mean squared observation error given by
    \begin{equation}\label{eq:h2}
        \begin{array}{*{20}{c}}
    {J = \left\| {\left[ {{{\tilde \Phi }_w}~~{{\tilde \Phi }_v}} \right]\left[ {\begin{array}{*{20}{c}}
    {\bm{K}}\\
    { - {\bm{I}}}
    \end{array}} \right]{\sigma _v} + {{\tilde \Phi }_w}\left[ {\Delta {\bm{A}}~~\Delta {\bm{L}}} \right]\left[ {\begin{array}{*{20}{c}}
    {{\Phi _A}{\bm{K}}}\\
    {\bm{I}}
    \end{array}} \right]{\sigma _v}} \right.}\\
    {{{\left. { + {{\tilde \Phi }_w}\left[ {\Delta {\bm{A}}~~\Delta {\bm{B}}} \right]\left[ {\begin{array}{*{20}{c}}
    {{\Phi _A}{\bm{B}}}\\
    {\bm{I}}
    \end{array}} \right]{\sigma_u}} \right\|}_{{{\mathcal{ H}}_2}}}},
    \end{array}
    \end{equation}
    where $\Delta \bm{L} = \bm{K} - \bm{L}$ is the difference between designed observer gain and optimal gain.
\end{lemma}
The input $\bm{u}$ and $\bm{v}$ both influence the observer error as shown in \eqref{x_tildex}.
When $\bm{L}$ is designed based on Theorem \ref{th:2}, the upper bound of the mean squared observation error is given in the following Theorem.
\begin{theorem}\label{th.3}
    Consider system \eqref{LTI-system} with unknown dynamics estimated by estimator \eqref{eq:SVR}. The estimation error satisfies $\|\Delta \bm{A}\|\leq\epsilon_A$, $\|\Delta \bm{B}\|\leq\epsilon_B$ and $\|\Delta \bm{L}\|\leq\epsilon_L$
     with probability at least $1-\delta$. The upper bound of the mean squared error $J$ is given by
    \begin{equation}\label{eq:J1}
         \begin{aligned}
        J&
        \leq \left \|\tilde{\Phi}_w\bm{K}- \tilde{\Phi}_v]\right \|_{\mathcal{H}_2}\sigma_v\\
        &+\sqrt{2}\epsilon_1\left \|\tilde{\Phi}_w\right\|_{\mathcal{H}_2}
        \left\|\begin{bmatrix} 
        \Phi_A\bm{B}\\
        \bm{I}
        \end{bmatrix} \right\|_{\mathcal{H}_\infty}\sigma_u\\
        &+
        \sqrt{2}\epsilon_2
        \left \|\tilde{\Phi}_w\right\|_{\mathcal{H}_2}
        \left\|\begin{bmatrix} 
        \Phi_A\bm{K}\\
        \bm{I}
        \end{bmatrix} \right\|_{\mathcal{H}_\infty}\sigma_v
        ,
    \end{aligned}
    \end{equation}
    where $\epsilon_1 = \max\left\{\epsilon_A,\epsilon_B\right\}$ and $\epsilon_2 = \max\left\{\epsilon_A,\epsilon_L\right\}$.
\end{theorem}
       	\begin{proof}
Please see Appendix F.
\end{proof} \begin{table*}[h]\label{Tabel:2}
        \centering
        \caption{RMSEs of estimations based on the OLS method and the proposed method with varying process noises for different systems}
\begin{tabular}{cp{1.5cm}cp{1.5cm}cp{1.5cm}cp{1.5cm}cp{1.5cm}cp{1.5cm}cp{1.5cm}cp{1.5cm}cp{1.5cm}cp{1.5cm}c}
\toprule[1.5pt]
                &     & \multicolumn{2}{c}{$\sigma_w=0.1$~~~~~~} & \multicolumn{2}{c}{$\sigma_w=1$~~~~~~} & \multicolumn{2}{c}{$\sigma_w=10$~~~~~~} \\\cline{3-8}
                &     & OLS    & Ours        & OLS     & Ours        & OLS     & Ours         \\ \hline
\multirow{2}{*}{\begin{tabular}[c]{@{}c@{}}open-loop stable\\ system \eqref{eq:stable-system}\\~\end{tabular}}                & ~~~~~$\tilde{\bm{A}}$   & 0.0261 & 0.0265  & 0.0328  & 0.0354 & 0.0541  & \bf{0.0484}   \\
  & ~~~~~$\tilde{\bm{B}}$   & 0.0029 & 0.0038  & 0.0211  & 0.0315  & 0.4278  & \bf{0.2695}   \\
                \midrule[0.75pt]
\multirow{2}{*}{\begin{tabular}[c]{@{}c@{}}open-loop unstable\\ system \eqref{eq:unstable-system}\\~\end{tabular}}                & ~~~~~$\tilde{\bm{A}}$   & 0.0225 & 0.0236 &  0.0166   & 0.0165  & 0.0171  & \bf{0.0168}       \\
  & ~~~~~$\tilde{\bm{B}}$   & 0.0019 & 0.0036  & 0.0374 & 0.0359  & 0.3774  & \bf{0.2506}   \\
                \bottomrule[1.5pt]
\end{tabular}
\end{table*}

The term $\epsilon_L$ is bounded since $\bm{A}$ and $\tilde{\bm{A}}$ are bounded.
When $\epsilon_L \leq \epsilon_A$ or $\epsilon_L \leq \epsilon_B$, the combination of the last two terms in \eqref{eq:J1} is upper bounded by
    \begin{equation*}
\sqrt{2}\epsilon_1\left \|\tilde{\Phi}_w\right\|_{\mathcal{H}_2}
        \left \{
        \left\|\begin{bmatrix} 
        \Phi_A\bm{B}\\
        \bm{I}
        \end{bmatrix} \right\|_{\mathcal{H}_\infty}\sigma_u
        +\left\|\begin{bmatrix} 
        \Phi_A\bm{K}\\
        \bm{I}
        \end{bmatrix} \right\|_{\mathcal{H}_\infty}\sigma_v
        \right \}.
    \end{equation*}
From Corrollary \ref{co:1}, we directly have that $\epsilon_{1} \sim \mathcal{O}\left(\frac{\gamma}{\sqrt{N}}\right)$, since $\epsilon_A\sim \mathcal{O}\left(\frac{\gamma}{\sqrt{N}}\right)$ and $\epsilon_B\sim \mathcal{O}\left(\frac{\gamma}{\sqrt{N}}\right)$.
Therefore, $J$ is bounded by $ \mathcal{O}\left(\frac{\gamma}{\sqrt{N}}||\tilde{\Phi}_w||_{\mathcal{H}_2}+\left \|\tilde{\Phi}_w\bm{K}- \tilde{\Phi}_v]\right \|_{\mathcal{H}_2}\sigma_v\right)$, which shows the sample length and parameter $\gamma$ in our estimator both influence the observer performance.

The backbone of minimizing $J$ is to find an optimal $\bm{L}$. An alternating simplified formulation is introduced by using robust SLS optimization technique:
\begin{equation*}
\begin{aligned}
    J_{opt}=\min_{\tilde{\Phi}_w,\tilde{\Phi}_v}~~&F \left \|\tilde{\Phi}_w\right \|_{\mathcal{H}_2}+\left\|\tilde{\Phi}_w\bm{K}-\tilde{\Phi}_v \right\|_{\mathcal{H}2},\\
    \text{s.t.}~~&\left\|\tilde{\Phi}_w\right\|_{\mathcal{H}_2}\leq\mathcal{C},\\
    &\tilde{\Phi}_w(z\bm{I}-\tilde{\bm{A}})-\tilde{\Phi}_v\bm{C} = \bm{I},\\
    &\tilde{\Phi}_w,\tilde{\Phi}_v \in \frac{1}{z}\mathcal{RH}_\infty,
\end{aligned}
\end{equation*}
where $\mathcal{C}$ is a regulation parameter to bound $\left\|\tilde{\Phi}_w\right\|_{\mathcal{H}_2}$, $F=\sqrt{2}\epsilon_1
        \left \{
        \left\|\begin{bmatrix} 
        \Phi_A\bm{B}\\
        \bm{I}
        \end{bmatrix} \right\|_{\mathcal{H}_\infty}\sigma_u
        +\left\|\begin{bmatrix} 
        \Phi_A\bm{K}\\
        \bm{I}
        \end{bmatrix} \right\|_{\mathcal{H}_\infty}\sigma_v
        \right \}$ is constant. The the affine constraints are used to parameterize the observer following the procedures in SLS\cite{wang2019system} and \cite{tsiamis2020sample}.
Further inspired by their work, we choose $\mathcal{C}$ as $\mathcal{C} \ge 2(1+\|\bm{K}\|)\|z\bm{I}-\bm{A}+\bm{KC}\|_{\mathcal{H}_2}$ and the estimation error satisfies $\epsilon_A\|\bm{I}-\bm{A}+\bm{KC}\|_{\mathcal{H}_\infty}\leq 1/2$.
Then, the optimal observation performance $J_{opt}$ fulfills
    \begin{equation*}\label{eq:J2}
         \begin{aligned}
        &J_{opt}
        \leq \epsilon_1  \{ \sqrt{2}\mathcal{C}
        \left \{
        \left\|\begin{bmatrix} 
        \Phi_A\bm{B}\\
        \bm{I}
        \end{bmatrix} \right\|_{\mathcal{H}_\infty }\sigma_u
        +\left\|\begin{bmatrix} 
        \Phi_A\bm{K}\\
        \bm{I}
        \end{bmatrix} \right\|_{\mathcal{H}_\infty }\sigma_v
        \right \} \\
        &+2\left\|z\bm{I}-\bm{A}+\bm{KC}\right\|_{\mathcal{H}_2}\sigma_v
        \}.
    \end{aligned}
    \end{equation*}
    This result shows that the performance of the optimal observer based on estimation with uncertainty follows $J_{opt}\sim \mathcal{O}\left(\frac{\gamma}{\sqrt{N}}\right)$. 
    When the system is fully observed, then $\sigma_v=0$ and $J_{opt}        \leq \sqrt{2} \epsilon_1   \mathcal{C}   \left\|\begin{bmatrix} 
        \Phi_A\bm{B}\\
        \bm{I}
        \end{bmatrix} \right\|_{\mathcal{H}_\infty }\sigma_u$, which means the observation only related to estimation error and $\left\|\tilde{\Phi}_w\right\|_{\mathcal{H}_2}$.
    The performance of a stable observer based on estimation with uncertainty is influenced by both $N$ and $\gamma$.
    By the merit of the adjustable parameter $\gamma$, the bias-variance trade-off is able to benefit the performance of the observation even with fixed number of samplings.
    
    \section{Numerical Simulations}\label{sec:6}

    In this section, several numerical simulations are conducted to illustrate the estimation and error bound based on the proposed method, as well as the optimality of the designed observer for an unknown LTI system. 
    We mainly focus on the influence of parameter $\gamma$.
    
    We consider an open-loop stable LTI system with state and input matrices
\begin{equation}\label{eq:stable-system}
    \bm{A} = \begin{bmatrix} 0.9 & 0.01 & 0\\ 0.01 & 0.9 & 0.01\\ 0 & 0.01 & 0.9 \end{bmatrix}, \bm{B} = \begin{bmatrix} 1\\ 1.5\\2 \end{bmatrix}.
\end{equation}
We also consider an open-loop unstable system, adapted from \cite{dean2020sample}, as follows
\begin{equation}\label{eq:unstable-system}
    \bm{A} = \begin{bmatrix} 1.01 & 0.01 & 0\\ 0.01 & 1.01 & 0.01\\ 0 & 0.01 & 1.01 \end{bmatrix}, \bm{B} = \begin{bmatrix} 1\\ 1.5\\2 \end{bmatrix}.
\end{equation}
In our experiments, we use stochastic input with $\sigma_u = 1$, test the performance of the method under different model noises with $\sigma_w = 0.1$, $\sigma_w = 1$ and $\sigma_w = 10$.
We assume that the state can be directly observed, i.e. $\bm{C} = \bm{I}$ for convenience.
For the multi-rollout setup, the rollout length is set to $T_0 = 11$, and we vary the number of rollouts from 10 to 450.
Empirically, $\gamma$ is often chosen to range in $(10^{-3},10^{-1})$ for the estimator.
Besides, due to the truncation error in computation, unsuitable $\gamma$ would lead to one of the terms in \eqref{eq:SVR-1} equals to 0, and cause the solution to fail.
We select $\gamma = 0.005$, $\gamma = 0.01$, $\gamma = 0.05$ and $\gamma = 0.1$.
LIBSVM is modified \cite{chang2011libsvm} with quadratic loss function.
For the error bound calculation, we select the matrices bound $M=1.1$ and possibility coefficient $\delta = 0.01$.
\subsection{Estimation for Systems with Varying Process Noise}
\begin{figure}[t]
    \subfigure[Estimation on open-loop stable system]
 {
  \begin{minipage}{8cm}
   \raggedleft
   \includegraphics[width=1\textwidth]{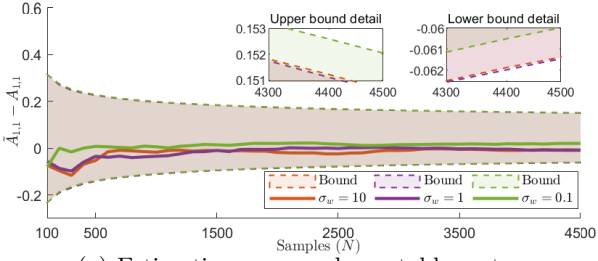}
  \end{minipage}
 }
    \subfigure[Estimation on open-loop unstable system]
    {
    \begin{minipage}{8cm}
   \raggedleft
   \includegraphics[width=1\textwidth]{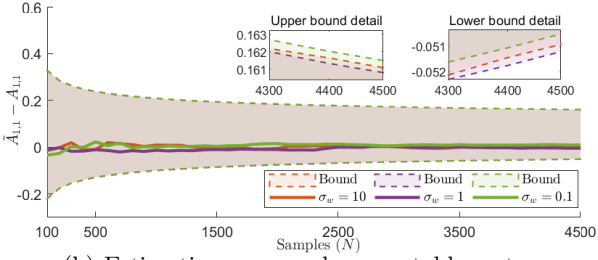}
  \end{minipage}
     }
     \caption{Estimation based on the proposed method for systems with varying process noise}
     \label{fig:different_noise}
\end{figure}
First, the estimator \eqref{eq:SVR} with $\gamma = 0.05$ is used to estimate the open-loop stable system \eqref{eq:stable-system} and open-loop unstable system \eqref{eq:unstable-system}.
The sample length used for estimation is set from $100$ to $4500$ to show the variation of error based on different sample numbers.
OLS is used to estimate the dynamics based on the same data for a fair comparison.
Root mean square error (RMSE) is used to measure the estimation error of $\bm{A}$, $\bm{B}$.
Table \ref{Tabel:2} shows The statistical results of 45 repeated estimations.
RMSE of the estimation of the proposed method is smaller than that of OLS when the process noise becomes larger, which is consistent with the theoretical analysis.
This illustrates that SVR is more suitable for dynamic estimation with large-variance noise since the lower variance of the error provides less volatile results.
The behavior of the proposed estimator with varying process noise is shown in Fig. \ref{fig:different_noise}.
The estimations of the proposed method for both stable and unstable systems fluctuate less slightly, thus showing more effectiveness.
The differences in estimation error on different process noises are small, which shows that the variance of process noise has little effect on the proposed estimator.
The error bounds given in \eqref{eq:18} are effective.
The error intervals tend to be smaller when introducing more samples to the estimation process.
Furthermore, the variance of process noise has little effect on error bounds, which shows the robustness of the proposed estimator \eqref{eq:SVR} and validate the error bound analysis \eqref{eq:18}.

The estimation error bounds given in \cite{dean2020sample,simchowitz2018learning,zheng2021sample,sun2020finite,oymak2021revisiting,sarkar2021finite,simchowitz2019learning} are not used for comparison, for our goal is to illustrate the influence of $\gamma$ on the error bound.

\subsection{Estimation with Varying Parameter of the Proposed Estimator}
\begin{figure}[t]
    \subfigure[Estimation on open-loop stable system]
 {
 \label{fig_3_1}
  \begin{minipage}{8cm}
   \raggedleft
   \includegraphics[width=1\textwidth]{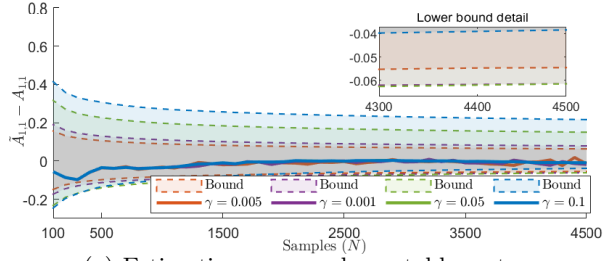}
  \end{minipage}
 }
    \subfigure[Estimation on open-loop unstable system]
    {
    \label{fig_3_2}
    \begin{minipage}{8cm}
   \raggedleft
   \includegraphics[width=1\textwidth]{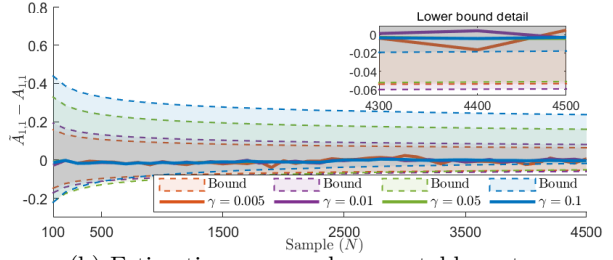}
  \end{minipage}
     }
     \caption{Estimation based on the proposed method with varying parameter}
     \label{fig:different_gamma}
\end{figure}
Then, the estimations of the proposed method with varying sample lengths and parameter $\gamma$ are conducted.
The behavior of the proposed estimator is shown in Fig. \ref{fig:different_gamma}.
The proposed estimator works well on both open-loop stable and unstable systems.
The error upper bounds of estimation significantly change with parameter $\gamma$.
This is due to the bias-variance trade-off in the proposed method.
It cannot be unbiased while holding low variance at the same time.
Moreover, as presented in the detail of the lower bound in Fig. \ref{fig_3_2}, the variation of the estimation is large when parameter $\gamma$ is small.
When $\gamma$ is small, the estimator seeks a smaller biased result, the upper bound is dragged to the x-axis, and the variance increases.
The variation of $\gamma$ also changes the interval between the upper and lower bound.
When seeking a small interval of the error bound, it can be done with smaller $\gamma$.
From the point of stable observer design, we want to have dynamics estimations that do not influence the stability's determination.
In Fig. \ref{fig_3_1}, $\bm{A}$ is determined to be stable after $N=1800$ for the error bound guarantees $\|\bm{A}\|<1$ as \eqref{eq:inter:A}.
Reducing the estimation error interval can be done by introducing more samples into the previous OLS-based estimation methods,
whereas the proposed estimator provides two ways (i.e., turning $\gamma$ and adding samples) to change the interval.



\begin{figure}[t]
    \centering
    \includegraphics[width=8cm]{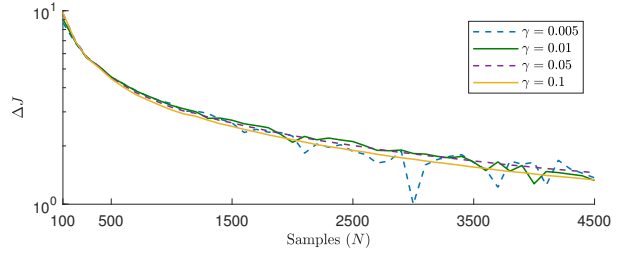}
    \caption{Deviation between observer cost bound and cost with different estimation parameter $\gamma$ on the open-looped stable system}
    \label{fig:deltaJ}
\end{figure}

\begin{figure}[t]
    \subfigure[Cost ratio between observers on open-loop stable system]
 {
  \begin{minipage}{8cm}
   \raggedleft
   \includegraphics[width=1\textwidth]{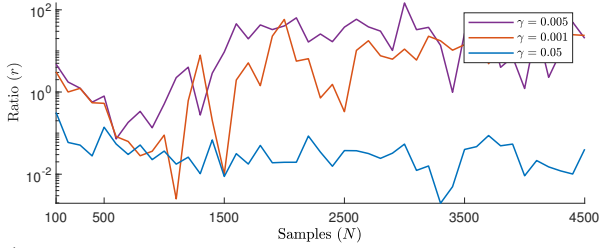}
  \end{minipage}
 }
    \subfigure[Cost ratio between observers on open-loop unstable system]
    {
    
    \begin{minipage}{8cm}
   \raggedleft
   \includegraphics[width=1\textwidth]{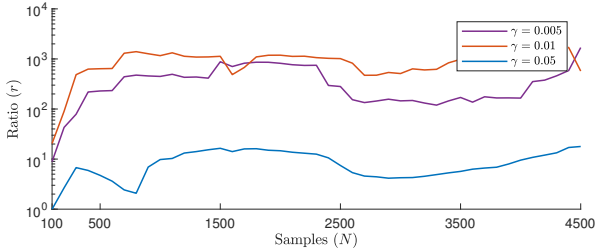}
  \end{minipage}
     }
     \caption{Ratios of observer costs with different estimation parameter $\gamma$,
     where $r:=|J_\gamma - J^*|/J^*$, $J_\gamma$ denotes the cost with different estimation parameter $\gamma$, $J^*$ is the benchmark cost with $\gamma = 0.1$.
     }
     \label{fig:cost}
\end{figure}
\subsection{Simulations on Observer Gain Design and Observer Performance Analysis}

Based on the estimation results with different $\gamma$, the stable observer gain is designed based on Theorem \ref{th:2}.
The stable gains are designed based on estimation with sample length from $100$ to $4500$.
We use the same gain if \eqref{eq:choose} holds.
The observer gain is designed easily on estimations with $\gamma = 0.1$, for the lower bound is closest to the x-axis after $N = 1000$ in Figure \ref{fig:different_gamma}.
Therefore, the cost $J$ is the most stable as $N$ increases when $\gamma = 0.1$.
The parameter $\gamma$ helps with the observer gain design.

Then, observers are constructed to verify the proposed cost bound and observer performance based on estimations with different $\gamma$.
We run 1000 Monte Carlo simulations for different sample lengths $N$.
The mean cost $J$ and cost bound $J_b$ for the observer of the open-loop stable system at each sample length is calculated.
The latter is obtained according to Theorem \ref{th.3} with $\epsilon_1 = \epsilon_B$, for $\epsilon_B$ is maximum among the candidates when $N \leq 4500$.
The deviation $\Delta J = J_b-J$ at each sample length is shown in Fig. \ref{fig:deltaJ}.
Deviations are above 0, which means the cost bound is valid with different $\gamma$.
Furthermore, costs with different $\gamma$ are compared in the open-loop stable and unstable systems.
In Fig. \ref{fig:cost}, we show the ratios of observer costs with different $\gamma$.
Ratio $r:=|J_\gamma - J^*|/J^*$, where $J_\gamma$ denotes the cost with different estimation parameter $\gamma$, $J^*$ is the benchmark cost with $\gamma = 0.1$.
The proposed observer design procedure provides stable observations for both open-loop stable and unstable systems.
It is also clear that the cost variance is much larger when $\gamma$ is smaller.
This corresponds to the high-variation estimation when $\gamma$ is small.
It should be noted that the estimation tends to be the same as that of OLS when $\gamma \to 0$.
Thus, we recommend using a relatively large $\gamma$ to have a stable estimation result for a stable observer performance.

In conclusion, the simulations demonstrate that the parameter $\gamma$ of the proposed estimator influences the estimation results and observer performance, and a suitable $\gamma$ benefits the observer design and stable performance.

    \section{Conclusion}\label{sec:7}
This paper mainly focused on SVR-based observer design and analysis for unknown linear systems.
We showed the detailed system dynamic estimation procedure, including data collection and the estimator's formulation. 
Furthermore, we analyzed and formulated the sample complexity bounds for estimation error of the proposed method as $\mathcal{O}(1/\sqrt{N})$ with an adjustable parameter $\gamma$, which provides another way to change the estimation error interval besides introducing more samples.
We also proposed an observer gain design procedure to guarantee stability based on the dynamic estimation results and the error bound.
The procedure reveals that $\gamma$ helps find a stable gain set by adjusting the estimation error interval.
We further analyze the end-to-end sample complexity for the sub-optimal observer for an unknown system.
We formulated a mean square observation error bound for estimation and observer design, connecting the estimation error and observer performance.
Finally, numerical simulations verify the proposed methods.
Simulations also illustrated that the parameter $\gamma$ of the proposed method influences the estimation results and observer performance, and a suitable $\gamma$ benefits the observer design and stable performance.

We would like to show the possible future directions:
\begin{itemize}
    \item The estimation upper and lower bounds are conservative. A possible direction is formulating a tighter bound by introducing other concentration inequality into the scaling procedure or the dual form of the SVR.
    \item The parameter $\gamma$ influences the performance variance in estimation and observer (controller) performance. It is possible to connect the variation with $\gamma$ over a high probability. 
    \item SVR provides the potential for estimating nonlinear systems. Sample complexity can extend to nonlinear systems by introducing nonlinear kernel functions into SVR.
\end{itemize}

\begin{ack}  
The authors would like to thank Yushan Li and Xiaoming Duan for their valuable comments and suggestions for this paper.
\end{ack}
\bibliographystyle{elsarticle-num}        
\bibliography{article.bib}           

\begin{thebibliography}{10}
\expandafter\ifx\csname url\endcsname\relax
  \def\url#1{\texttt{#1}}\fi
\expandafter\ifx\csname urlprefix\endcsname\relax\def\urlprefix{URL }\fi
\expandafter\ifx\csname href\endcsname\relax
  \def\href#1#2{#2} \def\path#1{#1}\fi

\bibitem{ljung1999system}
L.~Ljung, System identification, Wiley encyclopedia of electrical and
  electronics engineering (1999) 1--19.

\bibitem{abdalmoaty2019linear}
M.~R.-H. Abdalmoaty, H.~Hjalmarsson, Linear prediction error methods for
  stochastic nonlinear models, Automatica 105 (2019) 49--63.

\bibitem{young1970instrumental}
P.~C. Young, An instrumental variable method for real-time identification of a
  noisy process, Automatica 6~(2) (1970) 271--287.

\bibitem{chou1997subspace}
C.~T. Chou, M.~Verhaegen, Subspace algorithms for the identification of
  multivariable dynamic errors-in-variables models, Automatica 33~(10) (1997)
  1857--1869.

\bibitem{haber2014subspace}
A.~Haber, M.~Verhaegen, Subspace identification of large-scale interconnected
  systems, IEEE Transactions on Automatic Control 59~(10) (2014) 2754--2759.

\bibitem{yu2019subspace}
C.~Yu, J.~Chen, M.~Verhaegen, Subspace identification of individual systems in
  a large-scale heterogeneous network, Automatica 109 (2019) 108517.

\bibitem{inoue2019subspace}
M.~Inoue, Subspace identification with moment matching, Automatica 99 (2019)
  22--32.

\bibitem{luders1973adaptive}
G.~Luders, K.~Narendra, An adaptive observer and identifier for a linear
  system, IEEE Transactions on Automatic Control 18~(5) (1973) 496--499.

\bibitem{zhang2002adaptive}
Q.~Zhang, Adaptive observer for multiple-input-multiple-output ({MIMO}) linear
  time-varying systems, IEEE Transactions on Automatic Control 47~(3) (2002)
  525--529.

\bibitem{oliva2016fixed}
P.~Oliva-Fonseca, J.~G. Rueda-Escobedo, J.~A. Moreno, Fixed-time adaptive
  observer for linear time-invariant systems, in: IEEE Conference on Decision
  and Control, IEEE, 2016, pp. 1267--1272.

\bibitem{carleo2017solving}
G.~Carleo, M.~Troyer, Solving the quantum many-body problem with artificial
  neural networks, Science 355~(6325) (2017) 602--606.

\bibitem{cherkassky2004practical}
V.~Cherkassky, Y.~Ma, Practical selection of {SVM} parameters and noise
  estimation for {SVM} regression, Neural networks 17~(1) (2004) 113--126.

\bibitem{lewis2009reinforcement}
F.~L. Lewis, D.~Vrabie, Reinforcement learning and adaptive dynamic programming
  for feedback control, IEEE Circuits and Systems Magazine 9~(3) (2009) 32--50.

\bibitem{wang2009adaptive}
F.-Y. Wang, H.~Zhang, D.~Liu, Adaptive dynamic programming: An introduction,
  IEEE Computational Intelligence Magazine 4~(2) (2009) 39--47.

\bibitem{gao2016adaptive}
W.~Gao, Z.-P. Jiang, Adaptive optimal output regulation via output-feedback: an
  adaptive dynamic programing approach, in: IEEE Conference on Decision and
  Control, IEEE, 2016, pp. 5845--5850.

\bibitem{wei2019discrete}
Q.~Wei, R.~Song, Z.~Liao, B.~Li, F.~L. Lewis, Discrete-time impulsive adaptive
  dynamic programming, IEEE Transactions on Cybernetics (2019).

\bibitem{bauer2000analysis}
D.~Bauer, M.~Jansson, Analysis of the asymptotic properties of the {MOESP} type
  of subspace algorithms, Automatica 36~(4) (2000) 497--509.

\bibitem{ljung1998system}
L.~Ljung, System identification, in: Signal analysis and prediction, Springer,
  1998, pp. 163--173.

\bibitem{cherkassky1999model}
V.~Cherkassky, X.~Shao, F.~M. Mulier, V.~N. Vapnik, Model complexity control
  for regression using {VC} generalization bounds, IEEE transactions on Neural
  Networks 10~(5) (1999) 1075--1089.

\bibitem{chalimourda2004experimentally}
A.~Chalimourda, B.~Sch{\"o}lkopf, A.~J. Smola, Experimentally optimal $\nu$ in
  support vector regression for different noise models and parameter settings,
  Neural Networks 17~(1) (2004) 127--141.

\bibitem{tsiamis2020sample}
A.~Tsiamis, N.~Matni, G.~Pappas, Sample complexity of {K}alman filtering for
  unknown systems, in: Learning for Dynamics and Control, PMLR, 2020, pp.
  435--444.

\bibitem{dean2020sample}
S.~Dean, H.~Mania, N.~Matni, B.~Recht, S.~Tu, On the sample complexity of the
  linear quadratic regulator, Foundations of Computational Mathematics 20~(4)
  (2020) 633--679.

\bibitem{zheng2021sample}
Y.~Zheng, L.~Furieri, M.~Kamgarpour, N.~Li, Sample complexity of linear
  quadratic {G}aussian ({LQG}) control for output feedback systems, in:
  Learning for Dynamics and Control, PMLR, 2021, pp. 559--570.

\bibitem{simchowitz2018learning}
M.~Simchowitz, H.~Mania, S.~Tu, M.~I. Jordan, B.~Recht, Learning without
  mixing: {T}owards a sharp analysis of linear system identification, in:
  Conference On Learning Theory, PMLR, 2018, pp. 439--473.

\bibitem{sun2020finite}
Y.~Sun, S.~Oymak, M.~Fazel, Finite sample system identification: {Optimal}
  rates and the role of regularization, in: Learning for Dynamics and Control,
  PMLR, 2020, pp. 16--25.

\bibitem{oymak2021revisiting}
S.~Oymak, N.~Ozay, Revisiting {H}o-{K}alman based system identification:
  robustness and finite-sample analysis, IEEE Transactions on Automatic Control
  (2021).

\bibitem{sarkar2021finite}
T.~Sarkar, A.~Rakhlin, M.~A. Dahleh, Finite time {LTI} system identification
  (2021).

\bibitem{simchowitz2019learning}
M.~Simchowitz, R.~Boczar, B.~Recht, Learning linear dynamical systems with
  semi-parametric least squares, in: Conference on Learning Theory, PMLR, 2019,
  pp. 2714--2802.

\bibitem{tsiamis2021linear}
A.~Tsiamis, G.~J. Pappas, Linear systems can be hard to learn, arXiv preprint
  arXiv:2104.01120 (2021).

\bibitem{wang2019system}
Y.-S. Wang, N.~Matni, J.~C. Doyle, A system-level approach to controller
  synthesis, IEEE Transactions on Automatic Control 64~(10) (2019) 4079--4093.

\bibitem{furieri2019input}
L.~Furieri, Y.~Zheng, A.~Papachristodoulou, M.~Kamgarpour, An input--output
  parametrization of stabilizing controllers: {Amidst} {Youla} and system level
  synthesis, IEEE Control Systems Letters 3~(4) (2019) 1014--1019.

\bibitem{zheng2020equivalence}
Y.~Zheng, L.~Furieri, A.~Papachristodoulou, N.~Li, M.~Kamgarpour, On the
  equivalence of {Youla}, system-level, and input--output parameterizations,
  IEEE Transactions on Automatic Control 66~(1) (2020) 413--420.

\bibitem{mehta2019high}
P.~Mehta, M.~Bukov, C.-H. Wang, A.~G. Day, C.~Richardson, C.~K. Fisher, D.~J.
  Schwab, A high-bias, low-variance introduction to machine learning for
  physicists, Physics reports 810 (2019) 1--124.

\bibitem{geman1992neural}
S.~Geman, E.~Bienenstock, R.~Doursat, Neural networks and the bias/variance
  dilemma, Neural computation 4~(1) (1992) 1--58.

\bibitem{valentini2004bias}
G.~Valentini, T.~G. Dietterich, Bias-variance analysis of support vector
  machines for the development of {SVM}-based ensemble methods, Journal of
  Machine Learning Research 5~(Jul) (2004) 725--775.

\bibitem{willems2005note}
J.~C. Willems, P.~Rapisarda, I.~Markovsky, B.~L. De~Moor, A note on persistency
  of excitation, Systems \& Control Letters 54~(4) (2005) 325--329.

\bibitem{vapnik1999nature}
V.~Vapnik, The nature of statistical learning theory, Springer science \&
  business media, 1999.

\bibitem{wahba1999support}
G.~Wahba, et~al., Support vector machines, reproducing kernel hilbert spaces
  and the randomized gacv, Advances in Kernel Methods-Support Vector Learning 6
  (1999) 69--87.

\bibitem{saunders1998ridge}
C.~Saunders, A.~Gammerman, V.~Vovk, Ridge regression learning algorithm in dual
  variables (1998).

\bibitem{alazard2005introduction}
D.~Alazard, Introduction to {K}alman filtering, SUPAERO (2005).

\bibitem{balas1993mu}
G.~J. Balas, J.~C. Doyle, K.~Glover, A.~Packard, R.~Smith, $\mu$-analysis and
  synthesis toolbox, MUSYN Inc. and The MathWorks, Natick MA (1993).

\bibitem{hjartarson2015lpvtools}
A.~Hjartarson, P.~Seiler, A.~Packard, {LPVTools}: A toolbox for modeling,
  analysis, and synthesis of parameter varying control systems,
  IFAC-PapersOnLine 48~(26) (2015) 139--145.

\bibitem{doyle1989state}
J.~Doyle, K.~Glover, P.~Khargonekar, B.~Francis, State-space solutions to
  standard h/sub 2/and h/sub infinity/control problems, IEEE Transactions on
  Automatic Control 34~(8) (1989) 831--847.

\bibitem{scherer2001theory}
C.~Scherer, Theory of robust control, Delft University of Technology (2001)
  1--160.

\bibitem{kailath2000linear}
T.~Kailath, A.~H. Sayed, B.~Hassibi, Linear estimation, no. BOOK, Prentice
  Hall, 2000.

\bibitem{chang2011libsvm}
C.-C. Chang, C.-J. Lin, {LIBSVM}: a library for support vector machines, ACM
  transactions on intelligent systems and technology (TIST) 2~(3) (2011) 1--27.

\bibitem{rojo2004support}
J.~L. Rojo-{\'A}lvarez, M.~Mart{\'\i}nez-Ram{\'o}n, M.~de~Prado-Cumplido,
  A.~Art{\'e}s-Rodr{\'\i}guez, A.~R. Figueiras-Vidal, Support vector method for
  robust {ARMA} system identification, IEEE transactions on signal processing
  52~(1) (2004) 155--164.

\bibitem{lugosi2019sub}
G.~Lugosi, S.~Mendelson, Sub-{G}aussian estimators of the mean of a random
  vector, The annals of statistics 47~(2) (2019) 783--794.

\bibitem{vershynin2010introduction}
R.~Vershynin, Introduction to the non-asymptotic analysis of random matrices,
  arXiv preprint arXiv:1011.3027 (2010).

\end{thebibliography}
\textbf{Appendix}
\appendix
\section{The procedure of solving SVR}\label{app:1}
	The primal-dual or Lagrange functional for \eqref{eq:SVR} is obtained by introducing non-negative Lagrange multipliers $\lambda^+,\lambda^-,\alpha^+$ and $\alpha^-$.
	\begin{equation}
	\begin{aligned}
		    \mathcal{LD}_i&(\left [ {\bm{A}}, \;  {\bm{B}}\right]_i,\xi^+,\xi^-,\lambda^+,\lambda^-,\alpha^+,\alpha^-)\\
		    & = \frac{1}{2}\left\|\left[ {\bm{A}}, \;  {\bm{B}}\right]_i\right\|^2 + \frac{1}{2\gamma}\sum_{j=1}^N(\xi_j^{+2}+\xi_j^{-2}) \\
		    & - \sum_{j=1}^N (\lambda_j^+\xi_j^+ +  \lambda_j^-\xi_j^-)\\
		    & + \sum_{j=1}^N \alpha_j^+(f_i({\bm{z}_{k}^{j}}) - \left[\bm{A},\; \bm{B} \right]_i{\bm{z}_{k}^{j}} - \xi_j^+) \\
		    & + \sum_{j=1}^N \alpha_j^-(-f_i(\bm{z}_{k}^{j}) + \left[\bm{A},\; \bm{B} \right]_i{\bm{z}_{k}^{j}} - \xi_j^-)
	\end{aligned}
	\end{equation}
	Then the derivative with respect to Lagrange multipliers are
    \begin{equation}\label{eq:I1_mutiplier}
    \begin{aligned}
        &\frac{\partial\mathcal{LD}_i}{\partial \left [ {\bm{A}}, \;  {\bm{B}}\right]_i} = \left [ {\bm{A}}, \;  {\bm{B}}\right]_i - \sum_{j=1}^N(\alpha^+_j - \alpha_j^-)\bm{z}_{k}^{j} = 0 \\
        &\frac{\partial\mathcal{LD}_i}{\partial \xi_j^{+(-)}} = \frac{1}{\gamma}\xi_j^{+(-)} - \lambda_j^{+(-)}  - \alpha_j^{+(-)}  = 0\\
        &\frac{\partial\mathcal{LD}_i}{\partial \lambda_j^{+(-)}} = \sum \xi_j^{+(-)} \leq 0\\
        &\frac{\partial\mathcal{LD}_i}{\partial \alpha_j^+} = f_i(\bm{z}_{k}^{j}) - \left[\bm{A},\; \bm{B} \right]_i{\bm{z}_{k}^{j}} - \xi_{j}^+ \leq 0\\ 
        &\frac{\partial\mathcal{LD}_i}{\partial \alpha_j^-} = -f_i(\bm{z}_{k}^{j}) + \left[\bm{A},\; \bm{B} \right]_i{\bm{z}_{k}^{j}} - \xi_{j}^- \leq 0\\ 
    \end{aligned}
    \end{equation}
    Further, according to complementary conditions, $\forall j$,
\begin{equation}
\begin{aligned}
    \alpha_j^+(f_i({\bm{z}_{k}^{j}}) - \left[\bm{A},\; \bm{B} \right]_i{\bm{z}_{k}^{j}} - \xi_j^+) = 0,\\
    \alpha_j^-(-f_i(\bm{z}_{k}^{j}) + \left[\bm{A},\; \bm{B} \right]_i{\bm{z}_{k}^{j}} - \xi_j^-) = 0,\\
    \lambda_j^{+(-)}\xi_j^{+(-)} = 0.
\end{aligned}
\end{equation}
Take $\bm{\alpha}^+-\bm{\alpha}^- = \bm{\alpha} \ge 0$,
the dual form of the maximum optimization problem can be written as 
	\begin{equation}\label{eq:SVM-maxmin}
		     \mathcal{L}_D = - \frac{1}{2} \bm{\alpha} ^\top
		     \bm{z}_k^\top \bm{z}_k \bm{\alpha} 
		     +f_i({\bm{z}_k}) \bm{\alpha}.
	\end{equation}
The function approximation is 
\begin{equation*}
    \left[\tilde {\bm{A}},\; {\tilde{\bm{B}}} \right]_i = \bm{\alpha}^\top \bm{z}_k^\top.
\end{equation*}

\section{Proof of Lemma \ref{lemma:1}}\label{proof:lemma1}
First we give the relationship between regression results of SVR and OLS.
For the sake of clear expression, we use different notations in this section.
The standard linear relationship is given as $\bm{y}=\bm{a}^\top\bm{x}+\bm{b}$, where $\bm{b}\sim \mathcal{N}(0, \sigma\bm{I})$. More specifically, the linear operator $a$ in our problem is $A$. We use $a$ here for a general illustration.  The estimation of $\bm{a}$ is constructed based on $\bm{y}$ and $\bm{x}$.

To ease the notation, we use $\bm{w}$ and $\bm{v}$ to represent the results of SVR and OLS in this proof, respectively.
Note that \eqref{eq:SVR} is the formulation of SVR when it is without magnitude term.
The following relationship can be obtained by analysing  $\frac{\partial \mathcal{L}}{\partial \bm{w}} =0$ according to \cite{rojo2004support}.
\begin{equation*}
    \bm{w}_i = \frac{1}{1+\gamma}\bm{v}_i
\end{equation*}
This conclusion is also given in \cite{rojo2004support} by geometrical decomposition of SVR and OLS. Then, the expectation of $\bm{w}$ is given as 
\[\mathbb{E}(\bm{w}_i) = \mathbb{E}\left (\frac{1}{1+\gamma} \bm{v}_i \right) = \frac{1}{1+\gamma} \mathbb{E}(\bm{v}_i) = \frac{1}{1+\gamma} \bm{a}_i.
\]
Note that OLS is an unbiased estimator when the system with Gaussian excitation, thus, ${E}(v)$ equals to actual value. The covariance of $w$ is given as
\[
\mathbb{V}(\bm{v}_i) = \frac{1}{1+\gamma} (\mathbb{V}(\bm{v}_i) + \gamma \bm{a}_i^2).
\]
Recall the covariance of OLS estimation is
\[
\mathbb{V}(\bm{v}) = \sigma (\bm{x}\bm{x}^\top)^{-1}.
\]
For $\|\bm{A}\|\leq M$ in Assumption. 1, thus, $\|\bm{a}\|\leq M$ in the linear relationship and we have 
\[
\mathbb{V}(\bm{w}) \leq \frac{1}{1+\gamma} (\sigma (\bm{x}\bm{x}^\top)^{-1} + \gamma M^2\bm{I}).
\]
Lemma.1 stands.
It also reveals that the SVR provides a biased estimation, whereas the estimation covariance is smaller than that of OLS when $\mathbb{V}(\bm{v}_i) \ge \bm{a}_i^2$. 
\section{Proof of Theorem \ref{tho:sample-complexity}}
    According to (1.1) of \cite{lugosi2019sub}, when $\bm{X}$ follows a normal distribution with mean $\mu$ and covariance matrix $\Sigma$, then the sample mean $\Bar{\mu}$ is also normal with mean $\mu$ and covariance matrix $(1/N)\Sigma$, and therefore, for any $0<\delta<1$, with probability at least $1-\delta$
    \begin{equation*}
    \|\Bar{\mu} - \mu \|
    \leq \sqrt{\frac{\text{Tr}({\Sigma})}{N}} + \sqrt{\frac{2\lambda_{\max}\log(1/\delta)}{N}},
    \end{equation*}
    where $\lambda_{\max}$ denotes the largest eigenvalue of $\Sigma$.
    According to the distribution of estimation $\tilde{\bm{A}}$ given by \eqref{eq:distribution-A},
    the nonasymptotic bound of estimation error is given as
    \begin{equation}\label{eq:non-as-A}
    \begin{aligned}
        &\|\tilde{\bm{A}}_i - {E}(\tilde {\bm{A}}_i) \|\\
        &\leq \sqrt{\frac{\text{Tr}\left [ \bm{\sigma}^2_A (\bm{x}_{T_0-1}\bm{x}_{T_0-1}^\top)^{-1} \right ]+n\gamma M^2}{(1+\gamma)N_0}}\\
        &+ \sqrt{\frac{2\left\{\text{Tr}\left [ \bm{\sigma}^2_A (\bm{x}_{T_0-1}\bm{x}_{T_0-1}^\top)^{-1} \right ]+n\gamma M^2\right\}\log(1/\delta)}{(1+\gamma)N_0}}\\
        &\leq \sqrt{\frac{\theta_A +n\gamma M^2}{(1+\gamma)N_0}}+ \sqrt{\frac{2(\theta_A+n\gamma M^2)\log(1/\delta)}{(1+\gamma)N_0}}
    \end{aligned}
    \end{equation}
    where $\theta_A = n(mM\sigma^2_u + \sigma^2_w)\| (\bm{x}_{T_0-1}\bm{x}_{T_0-1}^\top)^{-1} \|$ and $N_0 = (T_0-1)N$.
    According to Corollary 5.35 of \cite{vershynin2010introduction} and Lemma 2.3 of \cite{dean2020sample}, the nonasymptotic bound of $\| (\bm{x}_{T_0-1}\bm{x}_{T_0-1}^\top)^{-1} \|$ is given as 
    \begin{equation}\label{eq:xxt}
        \| (\bm{x}_{T_0-1}\bm{x}_{T_0-1}^\top)^{-1} \| \le \frac{\|(vv^\top)^{-1}\|}{nM^{2T_0-1}\sigma^2_u + M^{2T_0-2} \sigma^2_w}
    \end{equation}
    where $v$ is a quantity follows a standard Gaussian distribution, and with probability at least $1-\delta$,
    \begin{equation}\label{eq:vvt}
    \begin{aligned}
        \|(vv^\top)^{-1}\|^{\frac{1}{2}}&\leq \frac{1}{\sqrt{N}+\sqrt{n}+\sqrt{2\log(1/\delta)}}\\
        &\leq \frac{2}{\sqrt{N}}
    \end{aligned}
    \end{equation}
    Combine \eqref{eq:xxt} and \eqref{eq:vvt}, it gets
    \begin{equation}
        \theta_A \leq \frac{4n(mM\sigma^2_u + \sigma^2_w)}{N(nM^{2T_0-1}\sigma^2_u + M^{2T_0-2} \sigma^2_w)}
    \end{equation}
    The upper bound is used in \eqref{eq:non-as-A} as \\$\theta_A = \frac{4n(mM\sigma^2_u + \sigma^2_w)}{N(nM^{2T_0-1}\sigma^2_u + M^{2T_0-2} \sigma^2_w)}$, which does not influence the inequality of \eqref{eq:non-as-A}.

    Note that $\tilde{\bm{A}}_i - {E}(\tilde {\bm{A}}_i) =  \frac{\gamma}{1+\gamma} {\tilde{\bm{A}}_i - \frac{1}{1+\gamma}\Delta {\bm{A}}_i}$, and $\|\tilde{\bm{A}}_i - {E}(\tilde {\bm{A}}_i) \| = \frac{1}{1+\gamma}\|\gamma {\tilde{\bm{A}}_i - \Delta {\bm{A}}_i} \|$. 
    Theorem \ref{tho:sample-complexity} holds for $\bm{A}$ estimation.
    
        Similarly, the nonasymptotic bound of estimation error of $\tilde {\bm{B}}_{i}$ is
    \begin{equation}
    \begin{aligned}
        &\|\tilde{\bm{B}}_i - {E}(\tilde {\bm{B}}_i) \|\\
        &\leq \sqrt{\frac{\theta_B +m\gamma M^2}{(1+\gamma)N_0}}+ \sqrt{\frac{2(\theta_B+m\gamma M^2)\log(1/\delta)}{(1+\gamma)N_0}}\\
    \end{aligned}
    \end{equation}
    where $N_0 = (T_0-1)N$, $\theta_B = m(nM^{2T_0-1}\sigma^2_u + (M^{2T_0-2}+1) \sigma^2_w)\\
    \| (\bm{u}_{T_0-1}\bm{u}_{T_0-1}^\top)^{-1} \|$ which has the upper bound as
    \begin{equation}
       \theta_B =  \frac{4m(nM^{2T_0-1}\sigma^2_u + (M^{2T_0-2}+1) \sigma^2_w)}{N\sigma^2_u}
    \end{equation}
    \section{Proof of Lemma \ref{lemma:interval}}
    It is easy to show that for each element in $\tilde{\bm{A}}_i$ and $\Delta {\bm{A}}_i$,
    $\|\gamma {\tilde{\bm{A}}_{i,j} - \Delta {\bm{A}}_{i,j}} \| \leq \|\gamma {\tilde{\bm{A}}_i - \Delta {\bm{A}}_i} \|$.\\
    Define $\mathcal{H}_A = \mathcal{H}(\gamma,\theta_A,n,N,M,\delta) = \sqrt{\frac{\theta_A +n\gamma M^2}{(1+\gamma)N_0}}+ \sqrt{\frac{2(\theta_A+n\gamma M^2)\log(1/\delta)}{(1+\gamma)N_0}}$.
    For each elements in $\tilde{\bm{A}}_i$, it has
    \begin{equation}\label{eq:inequation}
    \begin{aligned}
        &\Delta{\bm{A}}_{i,j}^2 -  2\gamma\tilde{\bm{A}}_{i,j}\Delta{\bm{A}}_{i,j}\\ 
        & -(1+\gamma)\mathcal{H}(\gamma,\theta_A,n,N,M,\delta) +\gamma^2 \tilde{\bm{A}}_{i,j}^2 \leq 0,
    \end{aligned}
    \end{equation}
    
    Since the coefficient of $\Delta{\bm{A}}_{i,j}^2$ is $1$, the parabola has a minimum point and opens upward. 
    The limit points of $\Delta{\Bar {\bm{A}}}_{i,j}$ are obtained by solving \eqref{eq:inequation} as
    \begin{equation}
    \begin{aligned}
        &\gamma\tilde{\bm{A}}_{i,j}
        \pm
        \frac{1}{2}\sqrt{4\gamma^2\tilde{\bm{A}}^2_{i,j}-4(\gamma^2\tilde{\bm{A}}_{i,j}^2-(1+\gamma)\mathcal{H}_A)}  \\
        =& \gamma\tilde{\bm{A}}_{i,j}
        \pm
        \sqrt{(1+\gamma)\mathcal{H}_A}
    \end{aligned}
    \end{equation}
    Then,
    \begin{equation*}
    \Delta {\bm{A}_{i,j}} \in 
    \mathcal{B}\left(\gamma\tilde{\bm{A}}_{i,j},\sqrt{(1+\gamma)\mathcal{H}_A} \right).
    \end{equation*}

    Define $\mathcal{H}_B = \mathcal{H}(\gamma,\theta_B,n,N,M,\delta) = \sqrt{\frac{\theta_B +m\gamma M^2}{(1+\gamma)N_0}}+ \sqrt{\frac{2(\theta_B+m\gamma M^2)\log(1/\delta)}{(1+\gamma)N_0}}$.
    The interval of $\Delta {\bm{B}_{i,j}}$ can be obtained based the similar manipulations of that of $\Delta {\bm{A}_{i,j}}$.
    \section{Proof of Theorem \ref{th:2}}
	The main idea of designing a stable observer gain is keep the eigenvalue less than 1.
	Here we adapt a conservative Gershgorin circle theorem for a general form of matrix $\bm{A}-\bm{LC}$.
	
	The center of the disc of each row is $\|{\bm{A}_{i,i}}- \{\bm{LC}\}_{i,j}\|$, which is in a unit circle.
	The largest radius of the disc is $\|(1+\gamma){\bm{A}_{i,j}}- {\bm{L}_{i,j}}\|+\sqrt{(1+\gamma)\mathcal{H}_A}$.
	The condition is obvious when the disc is in the unit circle.
	
	\section{Proof of Theorem \ref{th.3}}

	According to Lemma. \ref{lemma:3}, it has
	 \begin{equation}\label{eq:h2}
        \begin{array}{*{20}{c}}
    {J = \left\| {\left[ {{{\tilde \Phi }_w}~~{{\tilde \Phi }_v}} \right]\left[ {\begin{array}{*{20}{c}}
    {\bm{K}}\\
    { - {\bm{I}}}
    \end{array}} \right]{\sigma _v} + {{\tilde \Phi }_w}\left[ {\Delta {\bm{A}}~~\Delta {\bm{L}}} \right]\left[ {\begin{array}{*{20}{c}}
    {{\Phi _A}{\bm{K}}}\\
    {\bm{I}}
    \end{array}} \right]{\sigma _v}} \right.}\\
    {{{\left. { + {{\tilde \Phi }_w}\left[ {\Delta {\bm{A}}~~\Delta {\bm{B}}} \right]\left[ {\begin{array}{*{20}{c}}
    {{\Phi _A}{\bm{B}}}\\
    {\bm{I}}
    \end{array}} \right]{\sigma_u}} \right\|}_{{{\mathcal{ H}}_2}}}},
    \end{array}
    \end{equation}
    Inspired by Lemma 2 of \cite{tsiamis2020sample} and Proposition 3.5 of \cite{dean2020sample}, the inequality is
    \begin{equation}\label{eq:h2}
    \begin{aligned}
    J &\leq \left\| {\left[ {{{\tilde \Phi }_w}~~{{\tilde \Phi }_v}} \right]\left[ {\begin{array}{*{20}{c}}
    {\bm{K}}\\
    { - {\bm{I}}}
    \end{array}} \right]{\sigma _v}}\right\|_{\mathcal{ H}_2} \\
    & + \left\|{{\tilde \Phi }_w}\left[ {\Delta {\bm{A}}~~\Delta {\bm{B}}} \right]\left[ {\begin{array}{*{20}{c}}
    {{\Phi _A}{\bm{B}}}\\
    {\bm{I}}
    \end{array}} \right]{\sigma_u} \right\|_{{{\mathcal{ H}}_2}}\\
    &+ \left\|{{\tilde \Phi }_w}\left[ {\Delta {\bm{A}}~~\Delta {\bm{L}}} \right]\left[ {\begin{array}{*{20}{c}}
    {{\Phi _A}{\bm{K}}}\\
    {\bm{I}}
    \end{array}} \right]{\sigma _v} \right\|_{\mathcal{ H}_2}\\
    &\leq \left \|\tilde{\Phi}_w\bm{K}- \tilde{\Phi}_v]\right \|_{\mathcal{H}_2}\sigma_v\\
        &+\left \|{{\tilde \Phi }_w}\left[ {\Delta {\bm{A}}~~\Delta {\bm{B}}} \right]\right\|_{\mathcal{H}_2}
        \left\|\begin{bmatrix} 
        \Phi_A\bm{B}\\
        \bm{I}
        \end{bmatrix} \right\|_{\mathcal{H}_\infty}\sigma_u\\
        &+
        \left \|{{\tilde \Phi }_w}\left[ {\Delta {\bm{A}}~~\Delta {\bm{L}}} \right]\right\|_{\mathcal{H}_2}
        \left\|\begin{bmatrix} 
        \Phi_A\bm{K}\\
        \bm{I}
        \end{bmatrix} \right\|_{\mathcal{H}_\infty}\sigma_v
        .
    \end{aligned}
    \end{equation}
    Since $\|\Delta \bm{A}\|\leq\epsilon_A$, $\|\Delta \bm{B}\|\leq\epsilon_B$, $\|\Delta \bm{L}\|\leq\epsilon_L$, $\epsilon_1 = \max\left\{\epsilon_A,\epsilon_B\right\}$ and $\epsilon_2 = \max\left\{\epsilon_A,\epsilon_L\right\}$, it has
        \begin{equation}
         \begin{aligned}
        J&
        \leq \left \|\tilde{\Phi}_w\bm{K}- \tilde{\Phi}_v]\right \|_{\mathcal{H}_2}\sigma_v\\
        &+\sqrt{2}\epsilon_1\left \|\tilde{\Phi}_w\right\|_{\mathcal{H}_2}
        \left\|\begin{bmatrix} 
        \Phi_A\bm{B}\\
        \bm{I}
        \end{bmatrix} \right\|_{\mathcal{H}_\infty}\sigma_u\\
        &+
        \sqrt{2}\epsilon_2
        \left \|\tilde{\Phi}_w\right\|_{\mathcal{H}_2}
        \left\|\begin{bmatrix} 
        \Phi_A\bm{K}\\
        \bm{I}
        \end{bmatrix} \right\|_{\mathcal{H}_\infty}\sigma_v
        .
    \end{aligned}
    \end{equation}
\end{document}